\documentclass[11pt,english,amssymb,nofootinbib,superscriptaddress]{revtex4}
\usepackage{lmodern}

\usepackage[T1]{fontenc}
\usepackage[latin9]{inputenc}
\usepackage{color}
\usepackage{float}
\usepackage{amsmath}
\usepackage{amssymb}
\usepackage{graphicx}
\usepackage{esint}

\makeatletter

\providecommand{\tabularnewline}{\\}

\@ifundefined{textcolor}{}
{%
 \definecolor{BLACK}{gray}{0}
 \definecolor{WHITE}{gray}{1}
 \definecolor{RED}{rgb}{1,0,0}
 \definecolor{GREEN}{rgb}{0,1,0}
 \definecolor{BLUE}{rgb}{0,0,1}
 \definecolor{CYAN}{cmyk}{1,0,0,0}
 \definecolor{MAGENTA}{cmyk}{0,1,0,0}
 \definecolor{YELLOW}{cmyk}{0,0,1,0}
}

\@ifundefined{definecolor}{\usepackage{color}}{}
\usepackage{babel}

\usepackage{latexsym}\usepackage{bm}

\makeatother

\usepackage{babel}
\begin{document}

\title{Born-Infeld Gravity with a Unique Vacuum and a Massless Graviton}

\author{\.{I}brahim Güllü }

\email{ibrahimgullu2002@gmail.com}

\selectlanguage{english}%

\affiliation{Department of Physics,\\
 Middle East Technical University, 06800, Ankara, Turkey}

\author{Tahsin Ça\u{g}r\i{} \c{S}i\c{s}man}

\email{tahsin.c.sisman@gmail.com}

\selectlanguage{english}%

\affiliation{Department of Astronautical Engineering,\\
University of Turkish Aeronautical Association, 06790 Ankara, Turkey}

\author{Bayram Tekin}

\email{btekin@metu.edu.tr}

\selectlanguage{english}%

\affiliation{Department of Physics,\\
 Middle East Technical University, 06800 Ankara, Turkey}

\date{\today}
\begin{abstract}
We construct an $n$-dimensional Born-Infeld type gravity theory that
has the same properties as Einstein's gravity in terms of the vacuum
and particle content: Namely, the theory has a unique viable vacuum
(maximally symmetric solution) and a single massless unitary spin-2
graviton about this vacuum. The BI gravity, in some sense, is the
most natural, minimal generalization of Einstein's gravity with a
better UV behavior, and hence, is a potentially viable proposal for
low energy quantum gravity. The Gauss-Bonnet combination plays a non-trivial
role in the construction of the theory. As an extreme example, we
consider the infinite dimensional limit where an interesting exponential
gravity arises. 
\end{abstract}
\maketitle

\section{Introduction}

Recently in \cite{Gullu-4BI}, we have constructed a Born-Infeld type
(BI) gravity theory in the metric formulation with the Lagrangian
density $\mathcal{L}=\sqrt{\det\left(g_{\mu\nu}+\gamma A_{\mu\nu}\right)}$
in $3+1$-dimensions  that has the following properties:
\begin{enumerate}
\item The theory is minimal in the sense that the tensor $A_{\mu\nu}$ is
constructed from the powers of the Riemann tensor up to quadratic
order, but it does not have the derivatives of the Riemann tensor
as in its electrodynamics analogues with the Lagrangian density $\sqrt{\det\left(g_{\mu\nu}+b\, F_{\mu\nu}\right)}$.
It is important to note that linear theories of the form such as $\sqrt{\det\left(g_{\mu\nu}+\gamma R_{\mu\nu}\right)}$
do not yield unitary excitations about their maximally symmetric vacua,
and hence, one has to include at least the quadratic terms in the
curvature \cite{Deser_Gibbons}.
\item The theory reduces to the cosmological Einstein's theory at the lowest
order in the small curvature expansion about the flat space or the
(anti)-de Sitter {[}(A)dS{]}, and to the Einstein--Gauss-Bonnet (EGB)
theory at the quadratic order.
\item The theory describes only massless gravitons around its flat or (A)dS
vacuum at any finite (truncated) order in the curvature expansion
or as a full theory (namely, if all powers of curvature are included). 
\item The theory has a unique viable vacuum: namely, there is a single maximally
symmetric solution for which the massless spin-2 excitation about
this solution is unitary.
\end{enumerate}
In the current work, we extend the discussion to $n$-dimensional
spacetimes where $n\ge4$, and construct BI gravities that have the
above properties. {[}We have studied the $n=3$ case in \cite{Gullu-BINMG,Gullu-cfunc}
which already yields a nice theory at the linear level in the curvature
inside the determinant.{]} Some of the discussion in generic $n$-dimensions
is similar to the four dimensional case, but as we shall show, there
are nontrivial complications beyond four dimensions for a theory to
satisfy the above properties. $n=3+1$ dimension is highly special
in the sense that the requirement for the existence of a maximally
symmetric vacuum and the requirement for the theory be tree-level
unitary are equal only in this particular dimension but yield different
constraints in all other dimensions. 

General Relativity with its UV and IR problems is at best an effective
theory which is expected to be modified. The best scenario is that
there exists a quantum theory of gravity, perhaps a theory of strings,
from which one constructs low energy gravity theories at any desired
order in perturbation theory in powers of curvature which will be
of the form
\begin{equation}
I=\int d^{4}x\left\{ \frac{1}{\kappa}\left(R-2\Lambda_{0}\right)+\sum_{r=2}^{\infty}a_{r}\left({\rm Riem,{\rm Ric},{\rm R},\nabla{\rm Riem},\dots}\right)^{r}\right\} .\label{eq:Generic_grav}
\end{equation}
In reality, it is extremely complicated to compute the relevant terms
in this effective quantum gravity action from the microscopic theory
beyond several lower order terms. We, then here, suggest an alternative
bottom-up approach and construct effective quantum gravity actions
which have the good properties of the cosmological Einstein theory
noted above as well as a better UV behavior. Up to now, in the literature,
low energy quantum gravity actions have been constructed basically
on the principles that they be diffeomorphism invariant, ghost-free,
and sometimes supersymmetric. Diffeomorphism invariance is easy to
satisfy, and hence, does not much constrain the theory. Ghost-freedom
and supersymmetry are hard to satisfy conditions and so there are
only a few theories with low powers of curvature, $R^{2}$, $R^{3}$,
and at best $R^{4}$, that satisfy these constraints. Our point of
view here comes from the observation that cosmological Einstein's
theory has two more crucial properties: \emph{uniqueness of its maximally
symmetric vacuum }and\emph{ unitarity of its single massless graviton}
about this vacuum. Once, more powers of curvature are added to the
Einstein-Hilbert action, these two properties are immediately lost
\cite{Stelle}. Non-uniqueness of the maximally symmetric vacuum in
gravity is highly troublesome since each solution is a spacetime on
its own with different asymptotic structures and there would be no
way to choose one vacuum over the other \cite{BIuniD-short}. Therefore,
as a principle of constructing low energy quantum gravity theories,
besides the diffeomorphism invariance, we impose that the theory should
have a unique maximally symmetric vacuum about which the only excitation
is a unitary spin-2 graviton just like the cosmological Einstein's
theory. A priori, these conditions might appear tremendously difficult
to satisfy since an action of the form (\ref{eq:Generic_grav}) with
all possible terms at every order will yield practically intractable
expressions. Therefore, we will use the Born-Infeld construction which
limits the possible terms as well as fixes the arbitrary numerical
factors at each order. The fact that all the desired properties of
Einstein's theory can be kept intact in a Born-Infeld type theory
is quite remarkable. Especially, the fact that the theory has only
a non-ghost massless graviton about a unique vacuum is highly desirable. 

Construction of unitary ``minimal'' BI gravity turned out to be
highly non-trivial in four dimensions. For example, the Gauss-Bonnet
(GB) term, being a topological invariant, does not change the classical
equations of motion in four dimensions, plays a vital role in the
BI theory: without the GB term, one cannot build unitary actions of
the type described above.

The layout of the paper is as follows: in Section-II, we describe
the generic BI theory to be studied and briefly recapitulate the vacua
and the spectrum of the EGB theory, show that out of its two possible
vacua one of them is unstable due to a ghost massless graviton. In
Section-III, we study the vacuum and the particle spectrum about the
vacuum for the BI gravity. In Section-IV we discuss the unitarity
of BI gravity about the flat space which is also relevant for the
unitarity of the theory at $O\left(R^{2}\right)$ about its (A)dS
vacuum. In Section-V, we study the unitarity of the theory in (A)dS
backgrounds which requires calculating the effects of all powers of
curvature, we also construct the infinite dimensional BI gravity.
We relegate some of the technical parts to the Appendices. For example,
in Appendix C, we prove the uniqueness of the viable vacuum.

\section{Constructing the Born-Infeld Action}

The theory we shall consider is defined by the action 
\begin{equation}
I=\frac{2}{\kappa\gamma}\int d^{n}x\,\left[\sqrt{-\det\left(g_{\mu\nu}+\gamma A_{\mu\nu}\right)}-\left(\gamma\Lambda_{0}+1\right)\sqrt{-g}\right],\label{eq:Generic_BI}
\end{equation}
where $\gamma$ is a dimensionful parameter (the BI parameter). Defining
\begin{equation}
\mathcal{W}_{\mu\nu}\equiv C_{\mu\rho\alpha\beta}C_{\nu}^{\phantom{\nu}\rho\alpha\beta},\qquad\mathcal{W}\equiv g^{\mu\nu}\mathcal{W}_{\mu\nu}
\end{equation}
 where $C_{\mu\alpha\nu\beta}$ is the Weyl tensor, the most general
form of the two tensor $A_{\mu\nu}$ at quadratic order can be written
as 

\begin{align}
A_{\mu\nu}= & R_{\mu\nu}+\beta S_{\mu\nu}+\gamma\Bigg(a_{1}\mathcal{W}_{\mu\nu}+a_{2}C_{\mu\rho\nu\sigma}R^{\rho\sigma}+a_{3}R_{\mu\rho}R_{\nu}^{\rho}+a_{4}S_{\mu\rho}S_{\nu}^{\rho}\Bigg)\nonumber \\
 & +\frac{\gamma}{n}g_{\mu\nu}\Bigg(b_{1}\mathcal{W}+b_{2}R_{\rho\sigma}^{2}+b_{3}S_{\rho\sigma}^{2}\Bigg).\label{eq:Generic_Amn}
\end{align}
Here and from now on, for brevity we shall denote $R_{\rho\sigma}^{2}=R_{\rho\sigma}R^{\rho\sigma}$.
As mentioned in the Introduction, one has to include in $A_{\mu\nu}$
at least the quadratic terms to find unitary theories. Here, $S_{\mu\nu}$
is the traceless-Ricci tensor defined as $S_{\mu\nu}\equiv R_{\mu\nu}-\frac{1}{n}g_{\mu\nu}R$,
and the constants $\beta$, $a_{i}$, and $b_{i}$ are dimensionless.
Note that we have not included the term $R_{\mu\nu}S^{\mu\nu}$ since
it can be written as 
\begin{equation}
R_{\mu\rho}S_{\nu}^{\rho}=\frac{1}{2}R_{\mu\rho}R_{\nu}^{\rho}+\frac{1}{2}S_{\mu\rho}S_{\nu}^{\rho}-\frac{1}{2n}g_{\mu\nu}\left(R_{\rho\sigma}^{2}-S_{\rho\sigma}^{2}\right).
\end{equation}
Therefore, all possible linearly independent terms are included in
(\ref{eq:Generic_Amn}). In four dimensions, due to the identity $\mathcal{W}_{\mu\nu}=\frac{1}{4}g_{\mu\nu}\mathcal{W}$,
one can do away $a_{1}$, but this is not possible in generic $n$
dimensions. Here, we shall mostly work with the Weyl--traceless-Ricci--Ricci
(CSR) basis instead of the Riemann--Ricci--curvature-scalar (RRR)
basis. For the advantage of the CSR basis over the RRR basis in the
discussion of the spectrum around the constant curvature backgrounds
see \cite{Gullu-4BI}. Nevertheless, since it is sometimes needed
to work in the RRR basis, we give the non-trivial transformations
between these two bases in the Appendix \ref{sec:Conversions-Between-Bases}. 

It is clear that with 8 dimensionless parameters, the theory (\ref{eq:Generic_Amn})
is too general to be of much use. So, our task is to first constrain
some of these parameters by using physical arguments, which is the
main goal of this work. In addition, one should also worry about the
appearance of $\gamma$: it could be considered as a new parameter
of Nature that appears in quantum gravity; and hence, related to the
string tension or one could also use the Newton's constant instead
of $\gamma$ if one wants to keep only one single dimensionful parameter
in the theory. As discussed in \cite{Gullu-4BI}, this will be valid
as long as the curvature satisfies $R\ll1/\ell_{p}^{2}$ , where $\ell_{p}$
is the Planck length. 

Now, let us consider how to constrain the most general action (\ref{eq:Generic_BI}):
we first require that the theory has a unique maximally symmetric
vacuum about which the only excitation is a single massless unitary
spin-2 graviton. Since this issue is quite subtle, let us expound
on it: the theory in principle could have many maximally symmetric
solutions, but only one of them is viable in the sense that excitations
about the nonviable vacua are ghost-like while the viable vacuum has
the desired excitation of a unitary massless graviton. These two properties
as mentioned above, having a unique vacuum and a massless graviton,
are the properties of cosmological Einstein's theory; hence, one may
expect that at the free theory level, the general Born-Infeld gravity
should have the same properties as Einstein's theory. As we shall
show, this is actually a strong condition which cannot be satisfied.
But, the weaker condition that the BI theory has the same free-level
properties as the EGB theory can be satisfied without any phenomenological
difference in four dimensions. Beyond four dimensions, Newton's constant
receives corrections due to the cosmological constant. 

We will show that the above mentioned constraints reduces the number
of dimensionless parameters to four and the $A_{\mu\nu}$ tensor becomes
\begin{align}
A_{\mu\nu}= & R_{\mu\nu}+\beta S_{\mu\nu}+\gamma\left(a_{1}\mathcal{W}_{\mu\nu}+a_{2}C_{\mu\rho\nu\sigma}R^{\rho\sigma}+\frac{\beta+1}{4}R_{\mu\rho}R_{\nu}^{\rho}+a_{4}S_{\mu\rho}S_{\nu}^{\rho}\right)\nonumber \\
 & +\frac{\gamma}{n}g_{\mu\nu}\left[\left(\frac{\left(n-1\right)^{2}}{4\left(n-2\right)\left(n-3\right)}-a_{1}\right)\mathcal{W}-\frac{\beta}{4}R_{\rho\sigma}^{2}+\left(\frac{\beta\left(\beta+2\right)}{2}+\frac{n\left(4-3n\right)}{4\left(n-2\right)^{2}}-a_{4}\right)S_{\rho\sigma}^{2}\right].\label{eq:Main_result_for_Amn_0}
\end{align}
Hence, with this $A_{\mu\nu}$, the BI gravity given by the action
(\ref{eq:Generic_BI}) has a single massless unitary graviton about
its unique vacuum just like Einstein's theory. To determine the four
remaining dimensionless parameters which are not fixed by the condition
that the theory has a single unitary massless spin-2 graviton, further
conditions such as causality, supersymmetry, the existence of the
spherically symmetric solutions and cosmologically viable solutions
could be imposed. We will consider these in a separate work. In the
absence of further constraints, one can entertain the idea of obtaining
simpler theories by setting the constants to particular values. As
an example of such a minimal theory, let us consider (\ref{eq:Main_result_for_Amn_0})
and set $\beta=-1$, $a_{1}=0$, $a_{2}=0$, $a_{4}=0$, which yield
a theory without dimensionless parameters and $A_{\mu\nu}$ can be
recast as 
\begin{equation}
A_{\mu\nu}=\frac{g_{\mu\nu}}{n}\left(R+\gamma\frac{\left(n-1\right)^{2}}{4\left(n-2\right)\left(n-3\right)}\chi_{GB}-\gamma\frac{\left(n-2\right)}{4n}R^{2}\right),\label{eq:minimal_A_m_n}
\end{equation}
where the GB combination is defined as $\chi_{GB}=R_{\mu\nu\rho\sigma}^{2}-4R_{\mu\nu}^{2}+R^{2}$.
Therefore, the full Lagrangian density of this BI theory becomes 
\begin{equation}
\mathcal{L}=\frac{2}{\kappa\gamma}\left(\left(1+\frac{\gamma}{n}R+\frac{\gamma^{2}\left(n-1\right)^{2}}{4n\left(n-2\right)\left(n-3\right)}\chi_{GB}-\frac{\gamma^{2}\left(n-2\right)}{4n^{2}}R^{2}\right)^{\frac{n}{2}}-\lambda_{0}-1\right),\label{eq:minimal_lagrangian}
\end{equation}
where $\lambda_{0}\equiv\gamma\Lambda_{0}$ is the dimensionless bare
cosmological constant. In even dimensions one can get rid off the
square root and so this BI gravity becomes a specific $|\text{Riem}|^{n}$
theory. Its four dimensional version, which is a $|\text{Riem}|^{4}$
was given in \cite{Gullu-4BI}. 

Since the EGB theory will play a major role, let us briefly discuss
its properties here. In $n$ dimensions, the most general quadratic
action that describes \emph{only} massless spin-2 excitations around
its flat or AdS vacuum is the EGB theory with the Lagrangian 
\begin{align}
\kappa\mathcal{L} & =R-2\Lambda_{0}+\gamma\chi_{GB}.
\end{align}
Flat space is a vacuum for $\Lambda_{0}=0$ and there are generically
two (A)dS vacua with $\Lambda_{\pm}=-\frac{1}{4\kappa f}\left(1\pm\sqrt{1+8\kappa f\Lambda_{0}}\right)$,
where $f=\gamma\frac{\left(n-3\right)\left(n-4\right)}{\left(n-1\right)\left(n-2\right)}$.
We can rewrite the Lagrangian in terms of the Weyl tensor, the Ricci
tensor, and the traceless-Ricci tensor as 
\begin{align}
\kappa\mathcal{L} & =R-2\Lambda_{0}+\gamma\left[\mathcal{W}+\frac{\left(n-3\right)\left(n-2\right)}{\left(n-1\right)}\left(R_{\mu\nu}^{2}-\frac{n^{2}}{\left(n-2\right)^{2}}S_{\mu\nu}^{2}\right)\right],\label{GB_combination}
\end{align}
where we have used the identity 
\begin{equation}
\mathcal{W}=R_{\mu\nu\rho\sigma}^{2}-\frac{4}{n-2}R_{\mu\nu}^{2}+\frac{2}{\left(n-1\right)\left(n-2\right)}R^{2}.
\end{equation}
To understand the particle content of the EGB theory, one linearizes
the field equations about one of its (A)dS vacua to get 
\begin{equation}
\frac{1}{\kappa_{e}}\mathcal{G}_{\mu\nu}^{L}=0,\label{eq:linearized_EGB}
\end{equation}
where the effective Newton's constant is $\frac{1}{\kappa_{e}}=\frac{1}{\kappa}+4\Lambda f$,
and $\mathcal{G}_{\mu\nu}^{L}$ is the linearized Einstein tensor
which reduces to $\mathcal{G}_{\mu\nu}^{L}=-\frac{1}{2}\left(\bar{\square}-\frac{4\Lambda}{\left(n-1\right)\left(n-2\right)}\right)h_{\mu\nu}$
for the transverse-traceless perturbations, $h_{\mu\nu}\equiv g_{\mu\nu}-\bar{g}_{\mu\nu}$.
Therefore, (\ref{GB_combination}) describes a unitary massless spin-2
graviton as long as $\frac{1}{\kappa_{e}}>0$. Let us compare this
condition with the condition that there be a maximally symmetric solution.
For the latter, one needs 
\begin{equation}
1+8\kappa f\Lambda_{0}\ge0,
\end{equation}
for the former one has
\begin{equation}
\frac{1}{\kappa}+4\Lambda f>0.
\end{equation}
We assume $\kappa>0$, therefore once the value of $\Lambda$ is plugged,
the second condition yields
\begin{equation}
\mp\sqrt{1+8\kappa f\Lambda_{0}}>0,
\end{equation}
which is not possible for the $\Lambda_{+}$ branch; namely, massless
spin-2 excitation is ghostlike about this vacuum but unitary for the
$\Lambda_{-}$ vacuum. The lesson we learn from this exercise is that
not all vacua are stable or, in our terminology, viable which will
be the case for generic BI gravity that we shall discuss. Note that
the $\Lambda_{0}=0$ case was already noted in \cite{Boulware-String}.

\section{Vacuum and Spectrum of the BI Theory}

For generic gravity theories of the Born-Infeld type, it is a cumbersome
task to find the vacua and the particle spectrum about any of the
vacuum using the conventional techniques such as finding the field
equations and linearizing them. As we discussed in \cite{Gullu-UniBI,Gullu-4BI}
there are short-cuts to these computations. Let us briefly recall
these short-cuts here for the sake of completeness: Consider a generic
action of the form 
\begin{equation}
I=\int d^{n}x\,\sqrt{-g}\, f\left(R_{\alpha\beta}^{\mu\nu}\right),\label{eq:generic_action}
\end{equation}
where $f$ is a smooth function of its argument. {[}Note that the
function $f$ could depend on any arbitrary covariant derivative of
the Riemann tensor which will not alter the following discussion of
finding the vacuum, but it will of course change the discussion of
the particle spectrum about the vacuum.{]} In order to find the maximally
symmetric solutions to this theory one constructs the so called ``equivalent
linear action'' (ELA), $I_{\text{ELA}}$, that has the same vacua
as (\ref{eq:generic_action}): 
\begin{equation}
I_{\text{ELA}}=\frac{1}{\kappa_{l}}\int d^{n}x\,\sqrt{-g}\left(R-2\Lambda_{0,l}\right).\label{eq:action_ELA}
\end{equation}
Here, $l$ denotes the ELA values and one has 
\begin{equation}
\frac{1}{\kappa_{l}}=\zeta,\qquad\frac{\Lambda_{0,l}}{\kappa_{l}}=-\frac{1}{2}\bar{f}+\frac{n\Lambda}{n-2}\zeta,\label{eq:zeta_functions}
\end{equation}
where $\zeta$ is defined as 
\begin{equation}
\left[\frac{\partial f}{\partial R_{\rho\sigma}^{\lambda\nu}}\right]_{\bar{R}_{\rho\sigma}^{\mu\lambda}}R_{\rho\sigma}^{\lambda\nu}\equiv\zeta R,
\end{equation}
and the barred quantities are evaluated at the maximally symmetric
vacuum given as 
\begin{equation}
\bar{R}_{\rho\sigma}^{\mu\lambda}=\frac{2\Lambda}{\left(n-1\right)\left(n-2\right)}\left(\delta_{\rho}^{\mu}\delta_{\sigma}^{\lambda}-\delta_{\sigma}^{\mu}\delta_{\rho}^{\lambda}\right),
\end{equation}
for example, $\bar{f}\equiv f\left(\bar{R}_{\rho\sigma}^{\mu\lambda}\right)$.
From (\ref{eq:action_ELA}), one sets $\Lambda=\Lambda_{0,l}$ which
then reduces (\ref{eq:zeta_functions}) to a compact expression 
\begin{equation}
\Lambda=\frac{n-2}{4\zeta}\,\bar{f}.
\end{equation}
Solutions of this equation are the possible vacua of the theory. Once
the vacua are found, one can consider fluctuations about these vacua.
This amounts to finding the $O\left(h_{\mu\nu}^{2}\right)$ action
which is usually very complicated. Again, fortunately, there is a
similar short-cut method given in \cite{Hindawi} which relies on
finding an ``equivalent quadratic curvature action (EQCA)'' that
has the same propagator structure as (\ref{eq:generic_action}). Since
these matters are discussed at length in our previous paper \cite{Gullu-4BI},
in what follows we shall only quote the final results.

\subsection{Determining the Vacua of the BI Theory\label{sec:Determining-the-Vacuum}}

The equivalent linearized action of the BI theory given in (\ref{eq:Generic_BI})
is
\begin{equation}
\kappa\mathcal{L}_{\text{ELA}}=\frac{1}{\kappa_{l}}\left(R-\frac{2}{\gamma}\lambda_{0,l}\right),
\end{equation}
where we have defined a dimensionless Newton's constant $\kappa_{l}$
and a dimensionless cosmological parameter $\lambda_{0,l}\equiv\gamma\Lambda_{0,l}$
which can be found as 
\begin{equation}
\kappa_{l}=\frac{\bar{b}^{\frac{\left(2-n\right)}{2}}}{1+\frac{2\gamma\bar{R}}{n}c},\qquad\lambda_{0,l}=\kappa_{l}\left(1+\lambda_{0}-\bar{b}^{\frac{n}{2}}\right)+\frac{\gamma\bar{R}}{2},\qquad\bar{b}\equiv1+\frac{\gamma\bar{R}}{n}+\left(\frac{\gamma\bar{R}}{n}\right)^{2}c,\label{eq:Effective_ELA_params}
\end{equation}
here we have defined $c\equiv a_{3}+b_{2}$. Then, since the vacua
of $\mathcal{L}_{\text{ELA}}$ is given by the equation 
\begin{equation}
\lambda=\lambda_{0,l},
\end{equation}
with the use of $\gamma\bar{R}=\frac{2n\lambda}{n-2}$, one arrives
at the algebraic equation that determines the possible maximally symmetric
vacua
\begin{equation}
\left(\lambda_{0}+1\right)\left(cx^{2}+x+1\right){}^{\frac{2-n}{2}}+cx^{2}-1=0,\label{eq:vacuum_equation}
\end{equation}
where $x\equiv\frac{2\lambda}{n-2}$ and $\lambda_{0}\ne-1$. This
will be one of the main equations that we shall study in constraining
the theory. For generic $n$ dimensions, the equation cannot be solved
explicitly, but this is not required: we will show that there is a
unique solution consistent with the unitarity of the theory. Namely,
there is an interval for $\lambda_{0}$ which will yield a single
real $\lambda$ consistent with the unitarity of the theory. This
is by itself a rather remarkable result since a priori (\ref{eq:vacuum_equation})
could have many distinct real solutions which will correspond to possible
universes out of which ours cannot be identified on the basis of energy
comparison. The possible solutions of (\ref{eq:vacuum_equation})
is a somewhat technical analysis for which we devote Appendix-C to
it.

\section{Unitarity Around Flat Backgrounds}

In principle, BI gravity is valid in both weak and strong gravity
regimes. Once the BI action is considered as an expansion in curvature,
that is in $\gamma R$, depending on how well the inequality $\gamma R\ll1$
is satisfied, the relevant number of terms in the curvature expansion
of the BI theory changes. Thus, depending on the strength of gravitational
field under investigation, a truncated version of the curvature expansion
of the BI action can serve as an effective theory in that gravitational
regime. 

Now, considering BI theory is the main description of gravity, a natural
question is the unitarity of the excitations about the flat background.
To have such a unitarity analysis for the whole BI theory or its any
truncated order, one only needs to consider the terms up to $O\left(R^{2}\right)$
in the curvature expansion since the higher order terms do not have
any contribution to the free theory, that is the $O\left(h_{\mu\nu}^{2}\right)$
action about flat backgrounds. For the flat space $\lambda=0$ so
we must set $\lambda_{0}=0$ as is clear from (\ref{eq:vacuum_equation}).
Then, expanding (\ref{eq:Generic_BI}) up to $O\left(R^{2}\right)$
with the help of the Taylor series expansion 
\begin{equation}
\Big(\det\left(1+M\right)\Big)^{1/2}=1+\frac{1}{2}\text{Tr}M+\frac{1}{8}\left(\text{Tr}M\right)^{2}-\frac{1}{4}\text{Tr}\left(M^{2}\right)+O\left(M^{3}\right),
\end{equation}
one arrives at 

\begin{align}
\kappa\mathcal{L}_{O\left(R^{2}\right)}= & R+\gamma\left(a_{1}+b_{1}\right)\mathcal{W}+\gamma\left(a_{3}+b_{2}-\frac{1}{2}+\frac{n}{4}\right)R_{\mu\rho}^{2}\nonumber \\
 & +\gamma\left(a_{4}+b_{3}-\frac{n}{4}-\frac{\beta\left(\beta+2\right)}{2}\right)S_{\mu\rho}^{2}.\label{eq:flat_space_o(R^2)}
\end{align}
There are two possibilities now: one can either demand that the quadratic
terms vanish and one arrives at the Einstein theory or the quadratic
terms combine into the Gauss-Bonnet form yielding the EGB theory.
As we discussed before, these are the only theories that have massless
spin-2 gravitons and therefore both of these possibilities must be
separately analyzed. It will turn out that both of these possibilities
are viable as far as the flat space unitarity is concerned but in
what follows we will see that only the EGB reduction will yield a
viable theory when unitarity around the (A)dS space is studied. But,
that discussion requires the contributions of all the possible $O\left(R^{k}\right)$
terms with $k=1,2,3...,\infty$.

\subsection{Reduction to the Einstein theory}

We will reduce (\ref{eq:flat_space_o(R^2)}) to 
\begin{align}
\kappa\mathcal{L} & =R,
\end{align}
which requires the elimination of the quadratic terms which is possible
if the following identifications are made 
\begin{equation}
a_{1}=-b_{1},\qquad c=\frac{1}{2}-\frac{n}{4},\qquad a_{4}=\frac{\beta\left(\beta+2\right)}{2}+\frac{n}{4}-b_{3},\label{eq:flat_unitary_constraints}
\end{equation}
reducing the $A_{\mu\nu}$ tensor to a 5 parameter theory 
\begin{align}
A_{\mu\nu}= & R_{\mu\nu}+\beta S_{\mu\nu}+\gamma\Bigg(-b_{1}\mathcal{W}_{\mu\nu}+a_{2}C_{\mu\rho\nu\sigma}R^{\rho\sigma}\Bigg)\nonumber \\
 & +\gamma\left[\left(\frac{1}{2}-\frac{n}{4}-b_{2}\right)R_{\mu\rho}R_{\nu}^{\rho}+\left(\frac{\beta\left(\beta+2\right)}{2}+\frac{n}{4}-b_{3}\right)S_{\mu\rho}S_{\nu}^{\rho}\right]\nonumber \\
 & +\frac{\gamma}{n}g_{\mu\nu}\Bigg(b_{1}\mathcal{W}+b_{2}R_{\rho\sigma}^{2}+b_{3}S_{\rho\sigma}^{2}\Bigg).
\end{align}
Equation (\ref{eq:flat_unitary_constraints}) constitute the constraints
of the theory to be unitary about its flat vacuum. As expected these
constraints are not very restrictive. But as we mentioned above, if
one requires the $O\left(R^{2}\right)$ expansion of the BI theory
to be unitary around the (A)dS vacuum, then one has the same set of
constraints. Therefore, these constraints will be used later on.

\subsection{Reduction to the Einstein--Gauss-Bonnet theory}

We will compare the (\ref{eq:flat_space_o(R^2)}) with the EGB theory
\begin{align}
\kappa\mathcal{L} & =R+\gamma\left[\mathcal{W}+\frac{\left(n-3\right)\left(n-2\right)}{\left(n-1\right)}\left(R_{\mu\nu}^{2}-\frac{n^{2}}{\left(n-2\right)^{2}}S_{\mu\nu}^{2}\right)\right].\label{eq:GB}
\end{align}
Comparison of (\ref{eq:flat_space_o(R^2)}) and (\ref{eq:GB}) yield
the following identifications 
\begin{align}
c= & \frac{\left(n-2\right)\left(n-3\right)}{\left(n-1\right)}\left(a_{1}+b_{1}\right)+\frac{2-n}{4},\label{eq:c_equation}
\end{align}
\begin{align}
a_{4} & =\frac{\beta\left(\beta+2\right)}{2}-\frac{n^{2}\left(n-3\right)}{\left(n-1\right)\left(n-2\right)}\left(a_{1}+b_{1}\right)-b_{3}+\frac{n}{4},\label{eq:a_4_equation}
\end{align}
eliminating two of the parameters. Let us now turn to the thornier
issue of satisfying the tree-level unitarity around the (A)dS backgrounds.

\section{Unitarity Around $\text{(A)dS}$ Backgrounds}

In (A)dS backgrounds, unlike the flat space case infinitely many terms
contribute to the propagator and the free theory for the generic $n$-dimensional
BI gravity. Therefore, as explained above, we need the equivalent
quadratic curvature theory of
\begin{equation}
\kappa\mathcal{L}=\frac{2}{\gamma}\Bigg(\sqrt{\det\left(\delta_{\nu}^{\rho}+\gamma A_{\nu}^{\rho}\right)}-\left(\lambda_{0}+1\right)\Bigg),\label{eq:BI_action}
\end{equation}
which can be found after a Taylor series expansion whose details are
given in Appendix-B. After a long computation, one finally arrives
at 

\begin{equation}
\kappa\mathcal{L}_{\text{EQCA}}=\frac{1}{\tilde{\kappa}}\left(R-\frac{2}{\gamma}\tilde{\lambda}_{0}+\ensuremath{\alpha}_{1}\mathcal{W}+\ensuremath{\alpha}_{2}R_{\mu\nu}^{2}+\alpha_{3}S_{\mu\nu}^{2}\right).\label{eq:EQCA_of_BI}
\end{equation}
Let us again note the relation between (\ref{eq:BI_action}) and (\ref{eq:EQCA_of_BI}):
$O\left(h_{\mu\nu}\right)$ and $O\left(h_{\mu\nu}^{2}\right)$ expansions
of (\ref{eq:BI_action}) and (\ref{eq:EQCA_of_BI}) are identical
(they differ at $O\left(h_{\mu\nu}^{3}\right)$) granted that the
effective parameters, that are $\tilde{\kappa}$, $\tilde{\lambda}_{0}$,
$\alpha_{1}$, $\alpha_{2}$, $\alpha_{3}$, satisfy certain relations
as derived in Appendix-B and reproduced below. The dimensionless Newton's
constant and the dimensionless bare cosmological constant of the EQCA
theory are given as 
\begin{equation}
\frac{1}{\tilde{\kappa}}=\bar{b}^{\frac{\left(n-4\right)}{2}}\left(\bar{b}-\lambda\left(1+\frac{4\lambda}{n-2}c\right)^{2}\right),\label{eq:k_tilde_ads}
\end{equation}
\begin{equation}
\tilde{\lambda}_{0}=\tilde{\kappa}\left(\frac{n\lambda}{2\left(n-2\right)}\bar{b}^{\frac{\left(n-2\right)}{2}}\left(1+\frac{4\lambda}{n-2}c\right)+1-\bar{b}^{n/2}+\lambda_{0}\right)+\frac{n\lambda}{2\left(n-2\right)},\label{eq:l_0_tilde_ads}
\end{equation}
where with $\gamma\bar{R}=\frac{2n\lambda}{n-2}$, $\bar{b}$ of (\ref{eq:Effective_ELA_params})
simply takes the form 
\begin{equation}
\bar{b}=1+\frac{2\lambda}{n-2}\left(1+\frac{2\lambda}{n-2}c\right).\label{eq:bar_a}
\end{equation}
The coefficients of the quadratic parts are given as 
\begin{equation}
\alpha_{1}=\gamma\tilde{\kappa}\left(a_{1}+b_{1}\right)\bar{b}^{\frac{\left(n-2\right)}{2}},\qquad\alpha_{2}=\frac{\left(n-2\right)\gamma}{4\lambda}\left[\tilde{\kappa}\bar{b}^{\frac{\left(n-2\right)}{2}}\left(1+\frac{4\lambda}{n-2}c\right)-1\right],\label{eq:a1_and_a2_of_AdS}
\end{equation}
\begin{equation}
\alpha_{3}=\frac{\left(n-2\right)\gamma}{4\lambda}\left[\tilde{\kappa}\bar{b}^{\frac{\left(n-4\right)}{2}}\left(\bar{b}\left(\frac{4\lambda}{n-2}\left(a_{4}+b_{3}\right)-1\right)-\frac{2\lambda}{n-2}\left(\frac{4\lambda}{n-2}a_{3}+\beta+1\right)^{2}\right)+1\right],
\end{equation}
With the EQCA, (\ref{eq:EQCA_of_BI}), in our hands, we can now study
the unitarity of the BI gravity. Once again we have two options: we
can demand that this EQCA matches that of cosmological Einstein theory
or that of cosmological EGB theory.

\subsection{Reduction to the Einstein theory}

To reduce (\ref{eq:EQCA_of_BI}) to the cosmological Einstein theory
we must set the quadratic terms to zero: 
\begin{equation}
\alpha_{1}=\alpha_{2}=\alpha_{3}=0.
\end{equation}
 Recalling the constraints coming from the unitarity in flat backgrounds
(now these are equivalent to the unitarity of the $O\left(R^{2}\right)$
theory around (A)dS background) 
\begin{equation}
a_{1}=-b_{1},\qquad c=\frac{2-n}{4},\qquad a_{4}=\frac{\beta\left(\beta+2\right)}{2}+\frac{n}{4}-b_{3}.
\end{equation}
It is clear that $\alpha_{1}=0$ is automatically satisfied. On the
other hand, $\alpha_{2}=0$ gives another constraint on the parameters
of the theory as 
\begin{equation}
\frac{\left(n-2\right)\gamma}{4\lambda}\left[\tilde{\kappa}\bar{b}^{\frac{\left(n-2\right)}{2}}\left(1+\frac{4\lambda}{n-2}c\right)-1\right]=0.\label{eq:alpha_2}
\end{equation}
Since $c=\frac{2-n}{4}$, (\ref{eq:bar_a}) becomes 
\begin{equation}
\bar{b}=1+\frac{\lambda\left(2-\lambda\right)}{n-2}.
\end{equation}
Plugging this $\bar{b}$ in (\ref{eq:alpha_2}) yields $\lambda=2,$
or $\lambda=0$ \emph{independent of the number of dimensions}. Since
we already discussed the $\lambda=0$ case, let us consider the $\lambda=2$
case which yields $\bar{b}=1$ and $\tilde{\kappa}=-1$ which conflicts
the unitarity of the theory since the massless particle is a ghost.
This basically says that the BI theory cannot have vanishing quadratic
terms: hence, reduction to the unitary Einstein's theory is not possible
in any dimensions.

\subsection{Reduction to the Einstein--Gauss-Bonnet theory}

To simplify somewhat lengthy discussion, let us first recast the equivalent
quadratic action of the BI theory in the EGB form plus additional
quadratic curvature terms, after making use of the constraints coming
from the (A)dS unitarity of the theory at $O\left(R^{2}\right)$ which
boils down to the unitarity around the flat background. The conditions
are (\ref{eq:c_equation}) and (\ref{eq:a_4_equation}). In order
to investigate the theory in (A)dS backgrounds, our starting Lagrangian
is 

\begin{align}
\kappa\mathcal{L}_{\text{EQCA}}= & \frac{1}{\tilde{\kappa}}\left[R-\frac{2}{\gamma}\tilde{\lambda}_{0}+\text{\ensuremath{\alpha}}_{1}\left(\mathcal{W}+\frac{(n-3)}{(n-2)(n-1)}\left((n-2)^{2}R_{\mu\nu}^{2}-n^{2}S_{\mu\nu}^{2}\right)\right)+\text{\ensuremath{\tilde{\alpha}}}_{2}R_{\mu\nu}^{2}+\text{\ensuremath{\text{\ensuremath{\tilde{\alpha}}}_{3}}}S_{\mu\nu}^{2}\right],
\end{align}
where 
\begin{equation}
\text{\ensuremath{\tilde{\alpha}}}_{2}\equiv\ensuremath{\alpha}_{2}-\frac{\left(n-2\right)\left(n-3\right)}{\left(n-1\right)}\ensuremath{\alpha}_{1},\qquad\text{\ensuremath{\tilde{\alpha}}}_{3}\equiv\alpha_{3}+\frac{\left(n-3\right)n^{2}}{\left(n-2\right)\left(n-1\right)}\ensuremath{\alpha}_{1}.\label{eq:alpha_tildes}
\end{equation}
Once again unitarity is achieved by setting $\text{\ensuremath{\tilde{\alpha}}}_{2}=\text{\ensuremath{\tilde{\alpha}}}_{3}=0$
and $\tilde{\kappa}>0$. Using the constraints (\ref{eq:c_equation})
and (\ref{eq:a_4_equation}) $\tilde{\alpha}_{2}=0$ yields 
\begin{equation}
2\tilde{\kappa}\bar{b}^{\frac{\left(n-2\right)}{2}}\left(n-3\right)\left(a_{1}+b_{1}\right)=\frac{\left(n-1\right)}{2\lambda}\left(\tilde{\kappa}\bar{b}^{\frac{\left(n-2\right)}{2}}\left[\frac{\lambda\left(4\left(n-3\right)\left(a_{1}+b_{1}\right)-n+1\right)}{n-1}+1\right]-1\right).\label{eq:k_tilde_equation}
\end{equation}
Inserting $\tilde{\kappa}$ from (\ref{eq:k_tilde_ads}) (with the
assumption $1/\tilde{\kappa}\ne0$), (\ref{eq:k_tilde_equation})
reduces to 
\begin{align}
\left(a_{1}+b_{1}-\frac{\left(n-1\right)^{2}}{4\left(n-3\right)\left(n-2\right)}\right)\left(\frac{n-1}{2\left(n-3\right)}+\lambda\left(a_{1}+b_{1}-\frac{n-1}{4\left(n-3\right)}\right)\right)= & 0.\label{eq:a_1_and_b_1_cases}
\end{align}
The discussion bifurcates depending on the vanishing of the two factors.
We will study these below. The vanishing of $\tilde{\alpha}_{3}$
yields a complicated equation which we do not depict here.

\subsubsection{The two $a_{1}+b_{1}$ cases:}

As mentioned above, one has to discuss the two theories, coming from
the vanishing of the two factors in (\ref{eq:a_1_and_b_1_cases})
separately.

\paragraph{The $a_{1}+b_{1}\protect\ne\frac{\left(n-1\right)^{2}}{4\left(n-2\right)\left(n-3\right)}$
case:}

For this case, the second factor in (\ref{eq:a_1_and_b_1_cases})
vanishes yielding 
\begin{equation}
a_{1}+b_{1}=\frac{\left(\lambda-2\right)\left(n-1\right)}{4\lambda\left(n-3\right)}.\label{eq:First_a1+b1_value}
\end{equation}
With this result, (\ref{eq:c_equation}) reduces to 
\begin{equation}
c=\frac{2-n}{2\lambda},\label{eq:c_val_of_first_a1+b1_val}
\end{equation}
and the explicit forms of $\bar{b}$ and $\tilde{\kappa}$ can be
computed as 
\begin{align}
\bar{b} & =1,\qquad\frac{1}{\tilde{\kappa}}=1-\lambda,
\end{align}
where $\lambda\ne1$ and $\lambda\ne0$. First, observe that $x\equiv\frac{2\lambda}{n-2}=-\frac{1}{c}$;
therefore, $cx^{2}+x=0,$ which greatly simplifies the vacuum equation
(\ref{eq:vacuum_equation}) to
\begin{align}
\lambda_{0}+\frac{1}{c} & =0,
\end{align}
 fixing the cosmological parameter $\lambda$ as 
\begin{equation}
\lambda=\frac{\lambda_{0}}{2}\left(n-2\right).\label{eq:EoM_soln_for_first_a1+b1_val}
\end{equation}
Note that for $n=4$ the effective and the bare cosmological constants
are equal. In addition, with (\ref{eq:First_a1+b1_value}), the value
of $a_{4}+b_{3}$ can be calculated from (\ref{eq:a_4_equation})
as 
\begin{align}
a_{4}+b_{3} & =\frac{\beta\left(\beta+2\right)}{2}+\frac{n\left(n-\lambda\right)}{2\lambda\left(n-2\right)}.\label{eq:a_4_a_3_condition}
\end{align}
Plugging all these in (\ref{eq:alpha_tildes}) and setting $\tilde{\alpha}_{3}=0$
yields 
\begin{equation}
\lambda a_{3}\left(\frac{2\lambda a_{3}}{n-2}+\beta+1\right)=0.
\end{equation}
Once again one has to study the $a_{3}=0$ or $\frac{2\lambda a_{3}}{\left(n-2\right)}+\beta+1=0$
cases separately for $\lambda\ne0$.
\begin{enumerate}
\item The $a_{3}=0$ case: For $a_{3}=0$, $b_{2}$ can be determined from
(\ref{eq:c_val_of_first_a1+b1_val}) as 
\begin{align}
b_{2}= & \frac{2-n}{2\lambda}=c,
\end{align}
then the theory is fixed as
\begin{align}
A_{\mu\nu}= & R_{\mu\nu}+\beta S_{\mu\nu}+\gamma\left(a_{1}\mathcal{W}_{\mu\nu}+a_{2}C_{\mu\rho\nu\sigma}R^{\rho\sigma}+a_{4}S_{\mu\rho}S_{\nu}^{\rho}\right)\nonumber \\
 & +\frac{\gamma}{n}g_{\mu\nu}\Biggl[\left(\frac{\left(\left(n-2\right)\lambda_{0}-4\right)\left(n-1\right)}{4\lambda_{0}\left(n-2\right)\left(n-3\right)}-a_{1}\right)\mathcal{W}-\frac{1}{\lambda_{0}}R_{\rho\sigma}^{2}\nonumber \\
 & \phantom{+\frac{\gamma}{n}g_{\mu\nu}\Biggl[}+\left(\frac{\beta\left(\beta+2\right)}{2}+\frac{n\left(2n-\lambda_{0}\left(n-2\right)\right)}{2\lambda_{0}\left(n-2\right)^{2}}-a_{4}\right)S_{\rho\sigma}^{2}\Biggr],\label{eq:theory_1}
\end{align}
where we also used (\ref{eq:EoM_soln_for_first_a1+b1_val}) to represent
the action in terms of the input parameter $\lambda_{0}$ instead
of the derived parameter $\lambda$. With (\ref{eq:theory_1}), the
theory has four arbitrary dimensionless parameters which has all the
desired properties that Einstein's theory has.
\item The $a_{3}=-\frac{\left(n-2\right)}{2\lambda}\left(\beta+1\right)$
case: For this value $a_{3}$, $b_{2}$ is also determined from (\ref{eq:c_val_of_first_a1+b1_val})
as 
\begin{align}
b_{2}= & \frac{\left(n-2\right)}{2\lambda}\beta.
\end{align}
Then, the theory becomes 
\begin{align}
A_{\mu\nu}= & R_{\mu\nu}+\beta S_{\mu\nu}+\gamma\left[a_{1}\mathcal{W}_{\mu\nu}+a_{2}C_{\mu\rho\nu\sigma}R^{\rho\sigma}-\frac{\left(\beta+1\right)}{\lambda_{0}}R_{\mu\rho}R_{\nu}^{\rho}+a_{4}S_{\mu\rho}S_{\nu}^{\rho}\right]\nonumber \\
 & +\frac{\gamma}{n}g_{\mu\nu}\Biggl[\left(\frac{\left(\left(n-2\right)\lambda_{0}-4\right)\left(n-1\right)}{4\lambda_{0}\left(n-2\right)\left(n-3\right)}-a_{1}\right)\mathcal{W}+\frac{\beta}{\lambda_{0}}R_{\rho\sigma}^{2}\nonumber \\
 & \phantom{+\frac{\gamma}{n}g_{\mu\nu}\Biggl[}+\left(\frac{\beta\left(\beta+2\right)}{2}+\frac{n\left(2n-\lambda_{0}\left(n-2\right)\right)}{2\lambda_{0}\left(n-2\right)^{2}}-a_{4}\right)S_{\rho\sigma}^{2}\Biggr],\label{eq:theory_2}
\end{align}
where again we used (\ref{eq:EoM_soln_for_first_a1+b1_val}). This
is again a four parameter theory that has all the desired properties
of Einstein's theory. Even though both (\ref{eq:theory_1}) and (\ref{eq:theory_2})
provide a healthy extension of cosmological Einstein's theory with
a unique vacuum and a massless spin-2 graviton, they both lack the
$\lambda_{0}\rightarrow0$ limit. On the basis of this, we shall provisionally
disregard these two theories.
\end{enumerate}

\paragraph{The $a_{1}+b_{1}=\frac{\left(n-1\right)^{2}}{4\left(n-2\right)\left(n-3\right)}$
case:}

From (\ref{eq:c_equation}), this choice leads to
\begin{equation}
c=\frac{1}{4},\label{eq:c_val_of_second_a1+b1_val}
\end{equation}
and the explicit forms of $\bar{b}$ and $\tilde{\kappa}$ can be
computed as 
\begin{align}
\bar{b} & =\left(1+\frac{\lambda}{n-2}\right)^{2},
\end{align}
\begin{equation}
\frac{1}{\tilde{\kappa}}=\left(1-\lambda\right)\left(1+\frac{\lambda}{n-2}\right)^{n-2},\label{eq:kappa_tilde_of_the_theory}
\end{equation}
where $\lambda\ne1$ and $\lambda\ne2-n$ at which the effective dimensionless
Newton's constant vanish. With (\ref{eq:c_val_of_second_a1+b1_val}),
the vacuum equation reduces to a polynomial equation
\begin{equation}
\frac{\lambda}{n-2}-1+\left(\lambda_{0}+1\right)\left(\frac{\lambda}{n-2}+1\right)^{1-n}=0,\label{eq:Vacuum_of_the_theory}
\end{equation}
which has of course no explicit solutions for arbitrary $n$. But,
one can show that for given $n$, there is a unique viable solution
consistent with the unitarity of the theory (see Appendix-C). 

For this case, from (\ref{eq:a_4_a_3_condition}), $a_{4}+b_{3}$
becomes 
\begin{align}
a_{4}+b_{3} & =\frac{\beta\left(\beta+2\right)}{2}+\frac{n\left(4-3n\right)}{4\left(n-2\right)^{2}}.
\end{align}
Using this value in the second constraint of the Gauss-Bonnet reduction,
that is $\tilde{\alpha}_{3}=0$, one obtains 
\begin{align}
\left(\beta+1\right)\left(1+\frac{\lambda}{n-2}\right) & =\pm\left(\frac{4\lambda a_{3}}{n-2}+\beta+1\right).
\end{align}
One must consider both of the cases, but the minus sign case will
turn out to be a sub-case (when $\beta=-1$) of the plus sign case
for which
\begin{equation}
a_{3}=\frac{\beta+1}{4},
\end{equation}
and, $b_{2}=-\frac{\beta}{4}$. Then, the $A_{\mu\nu}$ tensor becomes
\begin{align}
A_{\mu\nu}= & R_{\mu\nu}+\beta S_{\mu\nu}+\gamma\left(a_{1}\mathcal{W}_{\mu\nu}+a_{2}C_{\mu\rho\nu\sigma}R^{\rho\sigma}+\frac{\beta+1}{4}R_{\mu\rho}R_{\nu}^{\rho}+a_{4}S_{\mu\rho}S_{\nu}^{\rho}\right)\nonumber \\
 & +\frac{\gamma}{n}g_{\mu\nu}\left[\left(\frac{\left(n-1\right)^{2}}{4\left(n-2\right)\left(n-3\right)}-a_{1}\right)\mathcal{W}-\frac{\beta}{4}R_{\rho\sigma}^{2}+\left(\frac{\beta\left(\beta+2\right)}{2}+\frac{n\left(4-3n\right)}{4\left(n-2\right)^{2}}-a_{4}\right)S_{\rho\sigma}^{2}\right].\label{eq:theory_4}
\end{align}
The BI gravity based on this $A_{\mu\nu}$, with four arbitrary dimensionless
parameters, satisfies all the nice properties of Einstein's theory:
a unique vacuum, a massless unitary spin-2 graviton about this vacuum.
In the small curvature expansion it reduces to the EGB theory while
with many powers of curvature it has an improved UV behavior. In Section-II,
we have discussed possible ways to further reduce the number of arbitrary
dimensionless parameters and suggested a possible minimal theory without
any such parameters given by the Lagrangian density (\ref{eq:minimal_lagrangian}).
For the sake of completeness let us note that the effective cosmological
parameter of the theory defined by (\ref{eq:theory_4}) will come
from the solution of (\ref{eq:Vacuum_of_the_theory}) consistent with
the positivity of the effective Newton's parameter (\ref{eq:kappa_tilde_of_the_theory})
whose details are in Appendix-C. Let us briefly summarize the results
of that analysis. Defining 
\begin{equation}
C\left(n\right)\equiv\left(\frac{n-3}{n-2}\right)\left(\frac{n-1}{n-2}\right)^{n-1}-1,
\end{equation}
one has the following conclusions. 
\begin{itemize}
\item \emph{For even dimensions there is a unique viable vacuum, $\lambda$,
in the region $-\infty<\lambda<1$ and $\lambda\ne2-n$ given that
$-\infty<\lambda_{0}<C\left(n\right)$ and $\lambda_{0}\ne-1$.}\textcolor{red}{{} }
\item \emph{For odd dimensions there is a unique viable vacuum, $\lambda$,
in the region $2-n<\lambda<1$ given that $-1<\lambda_{0}<C\left(n\right)$.}
\end{itemize}

\subsection{Infinite Dimensional BI Gravity:}

Without going into much detail, it is rather amusing to consider the
infinite dimensional ($n\rightarrow\infty$) limit, which received
a renewed interest in the context of $\frac{1}{n}$ expansion in general
relativity \cite{Emparan}. As $n\rightarrow\infty$, our minimal
BI Lagrangian (\ref{eq:minimal_lagrangian}) becomes an exponential
function compactly written as follows 
\begin{equation}
\kappa\frac{\gamma}{2}\mathcal{L}_{n\rightarrow\infty}=\exp\left(\frac{\gamma}{2}R+\frac{\gamma^{2}}{8}\left(R_{\mu\nu\rho\sigma}^{2}-4R_{\mu\nu}^{2}\right)\right)-\lambda_{0}-1.
\end{equation}
As we show in the Appendix-C this theory has a unique vacuum with
an effective cosmological parameter $\lambda=\ln(1+\lambda_{0})$
as long as $-1<\lambda_{0}<e-1$ and an effective Newton's constant
\begin{equation}
\frac{1}{\kappa_{\text{eff}}}=\frac{1}{\kappa}\Bigg(1-\ln\left(1+\lambda_{0}\right)\Bigg)\left(1+\lambda_{0}\right),
\end{equation}
which is always positive in the allowed region. Note that as $n\rightarrow\infty$
for (A)dS one has $\bar{R}_{\mu\nu\rho\sigma}=O\left(n^{-2}\right)$,
$\bar{R}_{\mu\nu}=O\left(n^{-1}\right)$, and $\gamma\bar{R}=2\lambda$.

\subsection{Conserved Charges in the BI Gravity:}

Finally let us briefly comment on the conserved charges (mass and
angular momenta) of asymptotically flat and (A)dS solutions in the
BI gravity. Since the conserved charges of the generic $f\left(R_{\alpha\beta}^{\mu\nu}\right)$
theory was given in \cite{Senturk} based on the formalism of \cite{Abbott-Deser,Deser_Tekin-PRL,Deser_Tekin}
with $\tilde{\kappa}$ given in (\ref{eq:kappa_tilde_of_the_theory})
it is straight forward to see that any conserved total charge of BI
gravity is given as 
\begin{equation}
Q_{\text{BI}}\left(\bar{\xi}\right)=\frac{1}{\tilde{\kappa}}\left(1+\frac{4\lambda\left(n-3\right)\left(n-4\right)}{\left(n-1\right)\left(n-2\right)}\alpha_{1}\right)Q_{\text{Einstein}}\left(\bar{\xi}\right),
\end{equation}
where $\tilde{\kappa}$ and $\alpha_{1}$ are given in (\ref{eq:k_tilde_ads})
and (\ref{eq:a1_and_a2_of_AdS}), respectively, and $\bar{\xi}$ is
the background Killing vector which reads $\bar{\xi}^{\mu}=\left(-1,0,...,0\right)$
for energy and $\bar{\xi}^{\mu}=\left(0,0,0,1,...,0\right)$ for angular
momenta. $Q_{\text{Einstein}}\left(\bar{\xi}\right)$ refers to the
charge of the solution in Einstein's theory. Hence, there is a simple
relation between the conserved charges of the BI gravity and Einstein's
theory. In particular, for asymptotically flat backgrounds they have
the same values. For asymptotically (A)dS backgrounds, they differ
a numerical factor depending on $\lambda$ and $n$. 

For the BI theory defined with (\ref{eq:theory_4}), the conserved
charge expression reads 
\[
Q_{\text{BI}}\left(\bar{\xi}\right)=\left(\frac{1}{\tilde{\kappa}}+\frac{\lambda\left(n-1\right)\left(n-4\right)}{\left(n-2\right)^{2}}\right)Q_{\text{Einstein}}\left(\bar{\xi}\right),
\]
where $\tilde{\kappa}$ is given in (\ref{eq:kappa_tilde_of_the_theory}).
It is interesting to note that for $n=4$, the second term drops out
and the BI theory has the same conserved charges as Einstein's theory.

\section{Conclusion}

Introducing the principle that the low energy quantum gravity has
a unique vacuum, that is a unique maximally symmetric solution, and
a single massless spin-2 graviton about this vacuum, we have constructed
Born-Infeld gravity theories in generic $n>3$ dimensions, including
$n\rightarrow\infty$. In $n$ dimensions the final theory has still
four arbitrary dimensionless parameters whose values could possibly
be determined from phenomenological considerations or other theoretical
conditions. The main motivation to construct such a theory was to
build a model which in principle has infinitely many terms in the
curvature invariants, hence improving Einstein's gravity in the UV
region, yet still has the two important properties of Einstein's gravity,
the uniqueness of the vacuum and the masslessness of the graviton,
which are usually lost when Einstein's theory is modified with higher
powers of curvature. Therefore, our construction answers the question
whether graviton can be kept massless in low energy quantum gravity
with a unique vacuum, in the affirmative. A detailed analysis of the
theory presented here in terms of its solutions will appear in a separate
work. It is interesting to note that $n=4$, the physically most relevant
case, has a rather fascinating property: BI gravity is not only unitary
as a full theory, but also unitary at every truncated order in the
curvature expansion \cite{Gullu-4BI}, hence in some sense one can
consider every truncated order as a separate theory in the strong
coupling limit. We are not aware of such a gravity theory. Whether
this property is preserved in higher dimensions needs to be studied.

Here, we have employed the metric formulation of BI gravity, for the
Palatini formulation, as envisioned by Eddington \cite{Eddington}
who introduced the determinantal type gravity based on generalized
volume a decade before Born and Infeld studied the electrodynamics
version \cite{BI}, see the recent works \cite{Banados_Eddington,Delsate_Steinhoff,Fiorini}.
See also \cite{Lavinia-Infrared,Lavinia-cascading} for various phenomenological
properties of other BI type gravities.

We have studied pure gravity without matter, to couple matter one
can consider the minimal coupling assumption and add a $\sqrt{-g}g^{\mu\nu}T_{\mu\nu}$
to the action where $T_{\mu\nu}$ is the usual energy-momentum tensor
of the matter fields. Of course, one can also use non-minimal coupling
such as $A_{\mu\nu}\rightarrow A_{\mu\nu}+\alpha F_{\mu\nu}+\eta\partial_{\mu}\phi\partial_{\nu}\phi$
where $F_{\mu\nu}$ is the field strength of electromagnetism and
$\phi$ is a scalar field.

\section{Acknowledgment}

I.~G. and B.~T. are supported by the TÜB\.{I}TAK grant 113F155.
T.~C.~S. thanks The Centro de Estudios Científicos (CECs) where
part of this work was carried out under the support of Fondecyt with
grant 3140127. Some of the calculations in this paper were either
done or checked with the help of the computer package Cadabra \cite{Cadabra-1,Cadabra-2}.

\appendix

\section{Conversions Between CSR Basis and RRR Basis\label{sec:Conversions-Between-Bases}}

In this Appendix, we discuss the conversions between the Weyl--traceless-Ricci--Ricci
(CSR) basis and the Riemann--Ricci--curvature-scalar (RRR) basis.
The $A_{\mu\nu}$ tensor written in the CSR basis, that is
\begin{align}
A_{\mu\nu}= & R_{\mu\nu}+\beta S_{\mu\nu}\nonumber \\
 & +\gamma\left(a_{1}C_{\mu\rho\sigma\lambda}C_{\nu}^{\phantom{\nu}\rho\sigma\lambda}+a_{2}C_{\mu\rho\nu\sigma}R^{\rho\sigma}+a_{3}R_{\mu\rho}R_{\nu}^{\rho}+a_{4}S_{\mu\rho}S_{\nu}^{\rho}\right)\nonumber \\
 & +\frac{\gamma}{n}g_{\mu\nu}\left(b_{1}C_{\rho\sigma\lambda\gamma}^{2}+b_{2}R_{\rho\sigma}^{2}+b_{3}S_{\rho\sigma}^{2}\right),
\end{align}
can be converted to the RRR basis, that is

\begin{align}
A_{\mu\nu}= & \left(1+\tilde{\beta}\right)R_{\mu\nu}-\frac{\tilde{\beta}}{4}g_{\mu\nu}R+c_{1}g_{\mu\nu}R^{2}+c_{2}RR_{\mu\nu}+c_{3}g_{\mu\nu}R_{\rho\sigma}^{2}\nonumber \\
 & +c_{4}R_{\phantom{\sigma}\mu}^{\sigma}R_{\nu\sigma}+c_{5}R_{\mu\sigma\nu\rho}R^{\sigma\rho}+c_{6}g_{\mu\nu}R_{\rho\sigma\lambda\gamma}^{2}+c_{7}R_{\mu}^{\phantom{\mu}\sigma\rho\tau}R_{\nu\sigma\rho\tau},\label{generic_A}
\end{align}
by using $S_{\mu\nu}=R_{\mu\nu}-\frac{1}{n}g_{\mu\nu}R$ and the definition
of the Weyl tensor in $n$ dimensions 
\begin{equation}
C_{\mu\alpha\nu\beta}=R_{\mu\alpha\nu\beta}-\frac{2}{\left(n-2\right)}\left(g_{\mu[\nu}R_{\beta]\alpha}-g_{\alpha[\nu}R_{\beta]\mu}\right)+\frac{2}{\left(n-1\right)\left(n-2\right)}Rg_{\mu[\nu}g_{\beta]\alpha}.
\end{equation}
The coefficients in (\ref{generic_A}) can be found as follows

\begin{align}
\tilde{\beta}= & \beta,\nonumber \\
c_{1}= & \frac{\gamma}{n^{2}\left(n-1\right)}\left(-\frac{2n^{2}}{\left(n-2\right)^{2}}a_{1}+\frac{n^{2}}{n-2}a_{2}+\left(n-1\right)a_{4}+\frac{2n}{n-2}b_{1}-\left(n-1\right)b_{3}\right),\nonumber \\
c_{2}= & \gamma\left(\frac{4}{\left(n-2\right)^{2}}a_{1}-\frac{n}{\left(n-1\right)\left(n-2\right)}a_{2}-\frac{2}{n}a_{4}\right),\nonumber \\
c_{3}= & \frac{\gamma}{n}\left(\frac{2n}{\left(n-2\right)^{2}}a_{1}-\frac{n}{n-2}a_{2}-\frac{4}{n-2}b_{1}+b_{2}+b_{3}\right),\nonumber \\
c_{4}= & \gamma\left(-\frac{2n}{\left(n-2\right)^{2}}a_{1}+\frac{2}{n-2}a_{2}+a_{3}+a_{4}\right),\nonumber \\
c_{5}= & \gamma\left(-\frac{4}{n-2}a_{1}+a_{2}\right),\quad c_{6}=\frac{\gamma}{n}b_{1},\quad c_{7}=\gamma a_{1}.\label{conversion_relations}
\end{align}
Sometimes the inverse transformation from the RRR basis to the CSR
basis is also needed; therefore, we shall give it here
\begin{align}
\beta= & \tilde{\beta},\quad a_{1}=\frac{c_{7}}{\gamma},\quad a_{2}=\frac{1}{\gamma}\left(c_{5}+\frac{4}{n-2}c_{7}\right),\nonumber \\
a_{3}= & \frac{1}{\gamma}\left(\frac{n}{2}c_{2}+c_{4}+\frac{n-2}{2\left(n-1\right)}c_{5}+\frac{2}{n-1}c_{7}\right),\nonumber \\
a_{4}= & \frac{1}{\gamma}\left(-\frac{n}{2}c_{2}-\frac{n^{2}}{2\left(n-1\right)\left(n-2\right)}c_{5}-\frac{2n}{\left(n-1\right)\left(n-2\right)^{2}}c_{7}\right),\nonumber \\
b_{1}= & \frac{nc_{6}}{\gamma},\quad b_{2}=\frac{1}{\gamma}\left(n^{2}c_{1}+\frac{n}{2}c_{2}+nc_{3}+\frac{n}{2\left(n-1\right)}c_{5}+\frac{2n}{n-1}c_{6}\right)\nonumber \\
b_{3}= & \frac{1}{\gamma}\left(-n^{2}c_{1}-\frac{n}{2}c_{2}+\frac{n^{2}}{2\left(n-1\right)\left(n-2\right)}c_{5}+\frac{2n^{2}}{\left(n-1\right)\left(n-2\right)}c_{6}+\frac{2n}{\left(n-2\right)^{2}}c_{7}\right).\label{Inverse_conversion_relations}
\end{align}

\section{Computation of the EQCA of BI in (A)dS }

Finding the vacuum and the particle spectrum of BI gravity is somewhat
tricky because of the contributions of all powers of curvature. Here,
basically we recap the essentials of the short-cuts introduced in
our earlier works \cite{Gullu-UniBI,Gullu-4BI,Sisman-AllUniD} as
applied to the present context. This boils down to finding an equivalent
quadratic curvature action (EQCA) that has the same vacuum and particle
spectrum as the BI gravity and that theory follows from the Taylor
series expansion. 

\begin{align}
\kappa\mathcal{L}_{\text{EQCA}}= & \frac{2}{\gamma}\left[\sqrt{\det\left(\delta_{\nu}^{\rho}+\gamma\bar{A}_{\nu}^{\rho}\right)}-\left(\lambda_{0}+1\right)\right]\nonumber \\
 & +\left[\frac{\partial\mathcal{L}}{\partial C_{\alpha\beta}^{\mu\nu}}\right]_{\bar{R}_{\rho\sigma}^{\mu\nu}}C_{\alpha\beta}^{\mu\nu}+\left[\frac{\partial\mathcal{L}}{\partial S_{\nu}^{\mu}}\right]_{\bar{R}_{\rho\sigma}^{\mu\nu}}S_{\nu}^{\mu}+\left[\frac{\partial\mathcal{L}}{\partial R_{\nu}^{\mu}}\right]_{\bar{R}_{\rho\sigma}^{\mu\nu}}\left(R_{\nu}^{\mu}-\bar{R}_{\nu}^{\mu}\right)\nonumber \\
 & +\frac{1}{2}\left[\frac{\partial^{2}\mathcal{L}}{\partial C_{\alpha\beta}^{\mu\nu}C_{\lambda\tau}^{\eta\theta}}\right]_{\bar{R}_{\rho\sigma}^{\mu\nu}}C_{\alpha\beta}^{\mu\nu}C_{\lambda\tau}^{\eta\theta}+\frac{1}{2}\left[\frac{\partial^{2}\mathcal{L}}{\partial S_{\nu}^{\mu}\partial S_{\beta}^{\alpha}}\right]_{\bar{R}_{\rho\sigma}^{\mu\nu}}S_{\nu}^{\mu}S_{\beta}^{\alpha}\nonumber \\
 & +\frac{1}{2}\left[\frac{\partial^{2}\mathcal{L}}{\partial R_{\nu}^{\mu}\partial R_{\beta}^{\alpha}}\right]_{\bar{R}_{\rho\sigma}^{\mu\nu}}\left(R_{\nu}^{\mu}-\bar{R}_{\nu}^{\mu}\right)\left(R_{\beta}^{\alpha}-\bar{R}_{\beta}^{\alpha}\right)\nonumber \\
 & +\left[\frac{\partial^{2}\mathcal{L}}{\partial C_{\alpha\beta}^{\mu\nu}\partial S_{\theta}^{\eta}}\right]_{\bar{R}_{\rho\sigma}^{\mu\nu}}C_{\alpha\beta}^{\mu\nu}S_{\theta}^{\eta}+\left[\frac{\partial^{2}\mathcal{L}}{\partial C_{\alpha\beta}^{\mu\nu}\partial R_{\theta}^{\eta}}\right]_{\bar{R}_{\rho\sigma}^{\mu\nu}}C_{\alpha\beta}^{\mu\nu}\left(R_{\theta}^{\eta}-\bar{R}_{\theta}^{\eta}\right)\nonumber \\
 & +\left[\frac{\partial^{2}\mathcal{L}}{\partial S_{\nu}^{\mu}\partial R_{\beta}^{\alpha}}\right]_{\bar{R}_{\rho\sigma}^{\mu\nu}}S_{\nu}^{\mu}\left(R_{\beta}^{\alpha}-\bar{R}_{\beta}^{\alpha}\right),\label{eq:Taylor_BI}
\end{align}
where the bracketed and barred quantities denote the maximally symmetric
background values for the corresponding expressions. Note that the
background values of the Weyl and the traceless Ricci scalar vanish
which was the main reason to work in the CRS basis. Let us compute
the terms of (\ref{eq:Taylor_BI}) separately. One can show that 
\begin{equation}
\left[\frac{\partial^{2}A_{\sigma}^{\rho}}{\partial C_{\alpha\beta}^{\mu\nu}\partial C_{\lambda\tau}^{\eta\theta}}\right]_{\bar{R}_{\rho\sigma}^{\mu\nu}}=\gamma a_{1}\delta_{\sigma}^{\alpha}\delta_{\theta}^{\beta}\delta_{\mu}^{\lambda}\delta_{\nu}^{\tau}\delta_{\eta}^{\rho}+\gamma\delta_{\eta}^{\alpha}\delta_{\theta}^{\beta}\delta_{\nu}^{\tau}\left(a_{1}\delta_{\sigma}^{\lambda}\delta_{\mu}^{\rho}+\frac{2b_{1}}{n}\delta_{\mu}^{\lambda}\delta_{\sigma}^{\rho}\right).
\end{equation}
 The other derivative terms can be calculated as 
\begin{equation}
\left[\frac{\partial^{2}A_{\sigma}^{\rho}}{\partial C_{\alpha\beta}^{\mu\nu}\partial R_{\theta}^{\eta}}\right]_{\bar{R}_{\rho\sigma}^{\mu\nu}}=\gamma a_{2}\delta_{\eta}^{\beta}\delta_{\nu}^{\theta}\delta_{\sigma}^{\alpha}\delta_{\mu}^{\rho},\qquad\left[\frac{\partial^{2}A_{\sigma}^{\rho}}{\partial S_{\nu}^{\mu}\partial S_{\beta}^{\alpha}}\right]_{\bar{R}_{\rho\sigma}^{\mu\nu}}=\gamma a_{4}\left(\delta_{\sigma}^{\nu}\delta_{\mu}^{\beta}\delta_{\alpha}^{\rho}+\delta_{\alpha}^{\nu}\delta_{\sigma}^{\beta}\delta_{\mu}^{\rho}\right)+\frac{2\gamma b_{3}}{n}\delta_{\alpha}^{\nu}\delta_{\mu}^{\beta}\delta_{\sigma}^{\rho},
\end{equation}
\begin{equation}
\left[\frac{\partial^{2}A_{\sigma}^{\rho}}{\partial R_{\nu}^{\mu}\partial R_{\beta}^{\alpha}}\right]_{\bar{R}_{\rho\sigma}^{\mu\nu}}=\gamma a_{3}\left(\delta_{\sigma}^{\nu}\delta_{\mu}^{\beta}\delta_{\alpha}^{\rho}+\delta_{\alpha}^{\nu}\delta_{\sigma}^{\beta}\delta_{\mu}^{\rho}\right)+\frac{2\gamma b_{2}}{n}\delta_{\alpha}^{\nu}\delta_{\mu}^{\beta}\delta_{\sigma}^{\rho}.
\end{equation}
It is also obvious that $\frac{\partial^{2}A_{\sigma}^{\rho}}{\partial C_{\alpha\beta}^{\mu\nu}\partial S_{\theta}^{\eta}}=0,\quad\frac{\partial^{2}A_{\sigma}^{\rho}}{\partial S_{\nu}^{\mu}\partial R_{\beta}^{\alpha}}=0.$
Using these results with 
\begin{equation}
\partial^{2}\left(\sqrt{\det\left(\delta_{\nu}^{\rho}+\gamma A_{\nu}^{\rho}\right)}\right)=\frac{\gamma}{2}\sqrt{\det\left(\delta_{\nu}^{\rho}+\gamma A_{\nu}^{\rho}\right)}\left[B_{\gamma}^{\lambda}\partial^{2}A_{\lambda}^{\gamma}-\gamma B_{\theta}^{\lambda}B_{\gamma}^{\tau}\left(\partial A_{\tau}^{\theta}\right)\partial A_{\lambda}^{\gamma}+\frac{\gamma}{2}\left(B_{\gamma}^{\lambda}\partial A_{\lambda}^{\gamma}\right)^{2}\right],
\end{equation}
where $B_{\gamma}^{\lambda}$ represents the inverse of the matrix
$\left(\delta_{\gamma}^{\lambda}+A_{\gamma}^{\lambda}\right)$ and
for the differential of $B$ we use $\partial B=-B\left(\partial A\right)B$.
The second order contributions to the EQCA can be expressed as
\begin{align}
\left[\frac{\partial^{2}\mathcal{L}}{\partial C_{\alpha\beta}^{\mu\nu}C_{\lambda\tau}^{\eta\theta}}\right]_{\bar{R}_{\rho\sigma}^{\mu\nu}}C_{\alpha\beta}^{\mu\nu}C_{\lambda\tau}^{\eta\theta}= & \sqrt{\det\left(\delta_{\nu}^{\mu}+\gamma\bar{A}_{\nu}^{\mu}\right)}\nonumber \\
 & \times\Biggl\{\bar{B}_{\rho}^{\sigma}\left[\frac{\partial^{2}A_{\sigma}^{\rho}}{\partial C_{\alpha\beta}^{\mu\nu}\partial C_{\lambda\tau}^{\eta\theta}}\right]_{\bar{R}_{\rho\sigma}^{\mu\nu}}-\gamma\bar{B}_{\zeta}^{\sigma}\left[\frac{\partial A_{\epsilon}^{\zeta}}{\partial C_{\lambda\tau}^{\eta\theta}}\right]_{\bar{R}_{\rho\sigma}^{\mu\nu}}\bar{B}_{\rho}^{\epsilon}\left[\frac{\partial A_{\sigma}^{\rho}}{\partial C_{\alpha\beta}^{\mu\nu}}\right]_{\bar{R}_{\rho\sigma}^{\mu\nu}}\nonumber \\
 & \phantom{x\Biggl\{}+\frac{\gamma}{2}\bar{B}_{\rho}^{\sigma}\left[\frac{\partial A_{\sigma}^{\rho}}{\partial C_{\alpha\beta}^{\mu\nu}}\right]_{\bar{R}_{\rho\sigma}^{\mu\nu}}\bar{B}_{\zeta}^{\epsilon}\left[\frac{\partial A_{\epsilon}^{\zeta}}{\partial C_{\lambda\tau}^{\eta\theta}}\right]_{\bar{R}_{\rho\sigma}^{\mu\nu}}\Biggr\} C_{\alpha\beta}^{\mu\nu}C_{\lambda\tau}^{\eta\theta},
\end{align}
which then yields 
\begin{equation}
\left[\frac{\partial^{2}\mathcal{L}}{\partial C_{\alpha\beta}^{\mu\nu}C_{\lambda\tau}^{\eta\theta}}\right]_{\bar{R}_{\rho\sigma}^{\mu\nu}}C_{\alpha\beta}^{\mu\nu}C_{\lambda\tau}^{\eta\theta}=\gamma^{2}\bar{b}^{\frac{\left(n-2\right)}{2}}\left(a_{1}+b_{1}\right)C_{\mu\nu\rho\sigma}^{2}.
\end{equation}
where $\bar{b}^{n}\equiv\det\left(\delta_{\nu}^{\beta}+\gamma\bar{A}_{\nu}^{\beta}\right)$.
Similarly, one has 
\begin{align}
\frac{\partial^{2}\mathcal{L}}{\partial S_{\nu}^{\mu}\partial S_{\beta}^{\alpha}}S_{\nu}^{\mu}S_{\beta}^{\alpha}= & \sqrt{\det\left(\delta_{\nu}^{\mu}+\gamma\bar{A}_{\nu}^{\mu}\right)}\nonumber \\
 & \times\Biggl\{\bar{B}_{\rho}^{\sigma}\left[\frac{\partial^{2}A_{\sigma}^{\rho}}{\partial S_{\nu}^{\mu}\partial S_{\beta}^{\alpha}}\right]_{\bar{R}_{\rho\sigma}^{\mu\nu}}-\gamma\bar{B}_{\zeta}^{\sigma}\left[\frac{\partial A_{\epsilon}^{\zeta}}{\partial S_{\beta}^{\alpha}}\right]_{\bar{R}_{\rho\sigma}^{\mu\nu}}\bar{B}_{\rho}^{\epsilon}\left[\frac{\partial A_{\sigma}^{\rho}}{\partial S_{\nu}^{\mu}}\right]_{\bar{R}_{\rho\sigma}^{\mu\nu}}\nonumber \\
 & \phantom{\times\Biggl[}+\frac{\gamma}{2}\bar{B}_{\rho}^{\sigma}\left[\frac{\partial A_{\sigma}^{\rho}}{\partial S_{\nu}^{\mu}}\right]_{\bar{R}_{\rho\sigma}^{\mu\nu}}\bar{B}_{\zeta}^{\epsilon}\left[\frac{\partial A_{\epsilon}^{\zeta}}{\partial S_{\beta}^{\alpha}}\right]_{\bar{R}_{\rho\sigma}^{\mu\nu}}\Biggr\} S_{\nu}^{\mu}S_{\beta}^{\alpha},
\end{align}
which yields 
\begin{equation}
\frac{\partial^{2}\mathcal{L}}{\partial S_{\nu}^{\mu}\partial S_{\beta}^{\alpha}}S_{\nu}^{\mu}S_{\beta}^{\alpha}=\gamma^{2}\bar{b}^{\frac{\left(n-4\right)}{2}}\Bigg(-\frac{1}{2}\beta^{2}+\bar{b}\left(a_{4}+b_{3}\right)\Bigg)S_{\mu\nu}^{2}.
\end{equation}
One also has
\begin{align}
\frac{\partial^{2}\mathcal{L}}{\partial R_{\nu}^{\mu}\partial R_{\beta}^{\alpha}}\left(R_{\nu}^{\mu}-\bar{R}_{\nu}^{\mu}\right)\left(R_{\beta}^{\alpha}-\bar{R}_{\beta}^{\alpha}\right)= & \sqrt{\det\left(\delta_{\nu}^{\mu}+\gamma\bar{A}_{\nu}^{\mu}\right)}\nonumber \\
 & \times\Biggl\{\bar{B}_{\rho}^{\sigma}\left[\frac{\partial^{2}A_{\sigma}^{\rho}}{\partial R_{\nu}^{\mu}\partial R_{\beta}^{\alpha}}\right]_{\bar{R}_{\rho\sigma}^{\mu\nu}}-\gamma\bar{B}_{\zeta}^{\sigma}\left[\frac{\partial A_{\epsilon}^{\zeta}}{\partial R_{\beta}^{\alpha}}\right]_{\bar{R}_{\rho\sigma}^{\mu\nu}}\bar{B}_{\rho}^{\epsilon}\left[\frac{\partial A_{\sigma}^{\rho}}{\partial R_{\nu}^{\mu}}\right]_{\bar{R}_{\rho\sigma}^{\mu\nu}}\nonumber \\
 & \phantom{\times\Biggl[}+\frac{\gamma}{2}\bar{B}_{\rho}^{\sigma}\left[\frac{\partial A_{\sigma}^{\rho}}{\partial R_{\nu}^{\mu}}\right]_{\bar{R}_{\rho\sigma}^{\mu\nu}}\bar{B}_{\zeta}^{\epsilon}\left[\frac{\partial A_{\epsilon}^{\zeta}}{\partial R_{\beta}^{\alpha}}\right]_{\bar{R}_{\rho\sigma}^{\mu\nu}}\Biggr\}\left(R_{\nu}^{\mu}-\bar{R}_{\nu}^{\mu}\right)\left(R_{\beta}^{\alpha}-\bar{R}_{\beta}^{\alpha}\right),
\end{align}
which reduces to 
\begin{align}
\frac{\partial^{2}\mathcal{L}}{\partial R_{\nu}^{\mu}\partial R_{\beta}^{\alpha}}\left(R_{\nu}^{\mu}-\bar{R}_{\nu}^{\mu}\right)\left(R_{\beta}^{\alpha}-\bar{R}_{\beta}^{\alpha}\right)= & -\left(R-\frac{\bar{R}}{2}-\frac{n}{2\bar{R}}R_{\mu\nu}^{2}+\frac{n}{2\bar{R}}S_{\mu\nu}^{2}\right)\nonumber \\
 & \phantom{-}\times\frac{\gamma^{2}\bar{R}}{2n^{3}}\bar{b}^{\frac{\left(n-4\right)}{2}}\left[4n^{2}\bar{b}\left(a_{3}+b_{2}\right)+(n-2)\left(2\gamma\bar{R}\left(a_{3}+b_{2}\right)+n\right)^{2}\right]\nonumber \\
 & -\frac{\gamma^{2}\bar{b}^{\frac{n}{2}-2}\left(\left(2a_{3}\gamma\bar{R}+n\right)^{2}-2n^{2}\bar{b}\left(a_{3}+b_{2}\right)\right)}{2n^{2}}S_{\mu\nu}^{2}
\end{align}
where we have used $R^{2}=n\left(R_{\mu\nu}^{2}-S_{\mu\nu}^{2}\right).$
Finally, the cross terms can be computed as 
\begin{equation}
\frac{\partial^{2}\mathcal{L}}{\partial C_{\alpha\beta}^{\mu\nu}\partial S_{\theta}^{\eta}}C_{\alpha\beta}^{\mu\nu}S_{\theta}^{\eta}=0,\qquad\frac{\partial^{2}\mathcal{L}}{\partial C_{\alpha\beta}^{\mu\nu}\partial R_{\theta}^{\eta}}C_{\alpha\beta}^{\mu\nu}\left(R_{\theta}^{\eta}-\bar{R}_{\theta}^{\eta}\right)=0,
\end{equation}
with only non-vanishing term coming from 
\begin{align}
\frac{\partial^{2}\mathcal{L}}{\partial S_{\nu}^{\mu}\partial R_{\beta}^{\alpha}}S_{\nu}^{\mu}\left(R_{\beta}^{\alpha}-\bar{R}_{\beta}^{\alpha}\right)= & \sqrt{\det\left(\delta_{\nu}^{\mu}+\gamma\bar{A}_{\nu}^{\mu}\right)}\nonumber \\
 & \times\Biggl\{\bar{B}_{\rho}^{\sigma}\left[\frac{\partial^{2}A_{\sigma}^{\rho}}{\partial S_{\nu}^{\mu}\partial R_{\beta}^{\alpha}}\right]_{\bar{R}_{\rho\sigma}^{\mu\nu}}-\gamma\bar{B}_{\zeta}^{\sigma}\left[\frac{\partial A_{\epsilon}^{\zeta}}{\partial R_{\beta}^{\alpha}}\right]_{\bar{R}_{\rho\sigma}^{\mu\nu}}\bar{B}_{\rho}^{\epsilon}\left[\frac{\partial A_{\sigma}^{\rho}}{\partial S_{\nu}^{\mu}}\right]_{\bar{R}_{\rho\sigma}^{\mu\nu}}\nonumber \\
 & \phantom{\times\Biggl[}+\frac{\gamma}{2}\bar{B}_{\rho}^{\sigma}\left[\frac{\partial A_{\sigma}^{\rho}}{\partial S_{\nu}^{\mu}}\right]_{\bar{R}_{\rho\sigma}^{\mu\nu}}\bar{B}_{\zeta}^{\epsilon}\left[\frac{\partial A_{\epsilon}^{\zeta}}{\partial R_{\beta}^{\alpha}}\right]_{\bar{R}_{\rho\sigma}^{\mu\nu}}\Biggr\} S_{\nu}^{\mu}\left(R_{\beta}^{\alpha}-\bar{R}_{\beta}^{\alpha}\right),
\end{align}
giving the result
\begin{equation}
\frac{\partial^{2}\mathcal{L}}{\partial S_{\nu}^{\mu}\partial R_{\beta}^{\alpha}}S_{\nu}^{\mu}\left(R_{\beta}^{\alpha}-\bar{R}_{\beta}^{\alpha}\right)=-\frac{1}{2}\gamma^{2}\bar{b}^{\frac{\left(n-4\right)}{2}}\beta\left(1+\frac{2\gamma\bar{R}a_{3}}{n}\right)S_{\mu\nu}^{2}.
\end{equation}

\section{Proof of the Uniqueness of the Viable Vacuum}

Now, for the theory (\ref{eq:theory_4}) let us discuss the viable
parameter regions (unitarity of the theory together with the existence
of a maximally symmetric vacuum). The discussion bifurcates for even
and odd dimensions $n$ which need to be studied separately. Before
the finite $n$ discussion, let us look at the extreme case of $n\rightarrow\infty$
limit which is relevant to the infinite dimensional BI gravity discussed
at the end of Section V. In this limit, (\ref{eq:kappa_tilde_of_the_theory})
becomes 
\begin{equation}
\frac{1}{\tilde{\kappa}}=\left(1-\lambda\right)e^{\lambda}.
\end{equation}
The positivity of $\tilde{\kappa}$, required for attractive gravity,
constrains the effective dimensionless cosmological constant to the
interval $-\infty<\lambda<1$. In this limit, the vacuum equation
(\ref{eq:Vacuum_of_the_theory}) becomes
\begin{equation}
\lambda_{0}=e^{\lambda}-1,\label{eq:lambda0_in_lambda_for_infinite_n-1}
\end{equation}
with the unique solution 
\begin{equation}
\lambda=\ln\left(1+\lambda_{0}\right).
\end{equation}
$\lambda$ is in the unitarity region, $\left(-\infty,1\right)$,
as long as the bare dimensionless cosmological constant satisfies
$-1<\lambda_{0}<e-1\approx1.7$, so there is a small interval for
$\lambda_{0}$.

Let us now turn to the discussion of the finite $n$ case. In analyzing
the existence of the roots of the vacuum equation (\ref{eq:Vacuum_of_the_theory}),
let us define a new variable 
\begin{equation}
y\equiv1+\frac{\lambda}{n-2},
\end{equation}
which reduces (\ref{eq:Vacuum_of_the_theory}) to
\begin{equation}
y^{n}-2y^{n-1}+a=0,\label{eq:vacuum_eq_with_y}
\end{equation}
where $a\equiv\lambda_{0}+1$. With this definition (\ref{eq:kappa_tilde_of_the_theory})
becomes 
\begin{equation}
\frac{1}{\tilde{\kappa}}=\left(n-1\right)y^{n-2}\left(1-\frac{n-2}{n-1}y\right).\label{eq:k_tilde_with_y}
\end{equation}
Our task is to prove that for generic $n$ dimensions (\ref{eq:vacuum_eq_with_y})
has at least one real solution consistent with the unitarity of the
theory. Surprisingly, it will turn out to be there is only one real
solution consistent with the unitarity. 

For a given $\lambda_{0}$, solving the algebraic equation (\ref{eq:vacuum_eq_with_y})
is a simple numerical problem for each given dimension $n$; but,
we do not know $\lambda_{0}$, hence, the problem becomes a non-trivial
one for $n\ge5$. The canonical way of showing the existence of the
roots in a given interval or finding approximate numerical solutions
is to construct the so called Sturm chain which we shall do below,
but to get a feeling let us analyze the function 
\begin{equation}
f\left(y\right)\equiv y^{n}-2y^{n-1}+a,
\end{equation}
whose zeros are the real solutions of the vacuum equation. To get
the information on the number of zeros, we need to study the extrema
of $f\left(y\right)$:
\begin{equation}
\frac{df}{dy}=ny^{n-2}\left(y-\frac{2\left(n-1\right)}{n}\right),
\end{equation}
which is zero at the two critical points $y=0$ and $y=\frac{2\left(n-1\right)}{n}$.
The second derivative of the function is
\begin{equation}
\frac{d^{2}f}{dy^{2}}=n\left(n-1\right)y^{n-3}\left(y-\frac{2\left(n-2\right)}{n}\right),
\end{equation}
which have the following values at the critical points:
\begin{equation}
\left.\frac{d^{2}f}{dy^{2}}\right|_{y=0}=0,\qquad\left.\frac{d^{2}f}{dy^{2}}\right|_{y=\frac{2\left(n-1\right)}{n}}=n\left(\frac{2\left(n-1\right)}{n}\right)^{n-2},
\end{equation}
showing that the first critical point is an inflection point and the
second one is a minimum. The value of the function at these points
are
\begin{equation}
f\left(0\right)=a,\qquad f\left(\frac{2\left(n-1\right)}{n}\right)=-\left(\frac{2}{n}\right)^{n}\left(n-1\right)^{n-1}+a.
\end{equation}
Clearly, depending on the signs of these values and the evenness or
the oddness of the number of dimensions, the number of roots can be
determined. 

Let us now construct the Sturm chain. The Sturm function is defined
as
\begin{equation}
f_{m}\left(x\right)=-\left(f_{m-2}\left(x\right)-f_{m-1}\left(x\right)\left[\frac{f_{m-2}\left(x\right)}{f_{m-1}\left(x\right)}\right]\right),
\end{equation}
where $\left[\frac{P\left(x\right)}{Q\left(x\right)}\right]$ is a
polynomial quotient. In other words, $f_{m}\left(x\right)$ is the
negative of the remainder in the polynomial division $\frac{f_{m-2}\left(x\right)}{f_{m-1}\left(x\right)}$,
that is $f_{m}\left(x\right)=-{\rm rem}\left(f_{m-2}\left(x\right),f_{m-1}\left(x\right)\right)$.
The zeroth and the first orders in the Sturm chain are the function
itself and its derivative:
\begin{align}
f_{0}\left(y\right) & =y^{n}-2y^{n-1}+a,\\
f_{1}\left(y\right) & =ny^{n-1}-2\left(n-1\right)y^{n-2}.
\end{align}
Then, $f_{2}\left(y\right)=-{\rm rem}\left(f_{0},f_{1}\right)$ and
$f_{3}\left(y\right)=-{\rm rem}\left(f_{1},f_{2}\right)$ can recursively
be calculated as
\begin{align}
f_{2}\left(y\right) & =\frac{4\left(n-1\right)}{n^{2}}y^{n-2}-a,\\
f_{3}\left(y\right) & =-\frac{n^{3}a}{4\left(n-1\right)}y+\frac{n^{2}a}{2}.
\end{align}
To find $f_{4}\left(y\right)$, one needs the quotient $\left[\frac{f_{2}\left(x\right)}{f_{3}\left(x\right)}\right]$
which can be computed as 
\begin{equation}
\left[\frac{f_{2}\left(x\right)}{f_{3}\left(x\right)}\right]=-\frac{4}{n^{3}a}y^{n-1}\sum_{i=2}^{n-1}\left[\frac{2\left(n-1\right)}{n\, y}\right]^{i},
\end{equation}
and $f_{4}\left(y\right)=-{\rm rem}\left(f_{2},f_{3}\right)$ becomes
\begin{equation}
f_{4}\left(y\right)=-\left(\frac{2}{n}\right)^{n}\left(n-1\right)^{n-1}+\lambda_{0}+1,
\end{equation}
which completes the Sturm chain. We can now use the Sturm Theorem
which reads (adapted to our notation) as \cite{Conk}:
\begin{quote}
The number of real roots of an algebraic equation ($f\left(y\right)=0$)
with real coefficients whose real roots are simple over an interval,
the endpoints of which are not roots, is equal to the difference between
the number of sign changes of the Sturm chains formed for the interval
ends. 
\end{quote}
To be able to use the Sturm Theorem, we need the relevant interval
where the theory is unitary. For this purpose even and odd dimensions
must be treated separately.

\subsubsection{Even dimensions: $n=4+2k$, $k=0,1,\dots$}

For unitarity $\tilde{\kappa}>0$ must be satisfied which requires
$\lambda<1$ and $\lambda\ne2-n$ in even dimensions as is clear from
(\ref{eq:k_tilde_with_y}). Therefore, for even dimensions, the relevant
interval is $y\in\left(-\infty,\frac{n-1}{n-2}\right)-\{0\}$. There
is a corresponding viable interval of $\lambda_{0}$ and to find this,
let us show that
\begin{equation}
\lambda_{0}=2y^{n-1}-y^{n}-1,\label{eq:lamda0_in_lambda-1}
\end{equation}
is a monotonically increasing function of $y$ with a positive derivative
in the unitarity region:
\begin{align}
\frac{d\lambda_{0}}{dy} & =-ny^{n-2}\left(y-\frac{2\left(n-1\right)}{n}\right).\label{eq:Derv_of_lambda0-1}
\end{align}
For even $n$, $d\lambda_{0}/dy$ changes sign at $y=\frac{2\left(n-1\right)}{n}$
which is not attained below the upper bound $y=\frac{n-1}{n-2}$ for
$n\ge4$; hence, $\frac{d\lambda_{0}}{dy}>0$. The upper bound of
$y$, that is $\lambda=1$, gives the corresponding upper bound for
$\lambda_{0}$ as
\begin{equation}
\lambda_{0}<C\left(n\right)\equiv\left(\frac{n-3}{n-2}\right)\left(\frac{n-1}{n-2}\right)^{n-1}-1.\label{eq:Unitarity_bound-1}
\end{equation}
The upper bound $C\left(n\right)$ is an increasing sequence since
$C\left(n\right)$ involves the multiplication of two increasing sequences:
\begin{equation}
\left(\frac{n-3}{n-2}\right)\left(\frac{n-1}{n-2}\right)=1-\frac{1}{\left(n-2\right)^{2}},
\end{equation}
and
\begin{equation}
\left(1+\frac{1}{n-2}\right)^{n-2},
\end{equation}
which converges to $e$ as $n\rightarrow\infty$. Therefore, as $n\rightarrow\infty$,
$C\left(n\right)$ converges to $e-1$.

\begin{table}[H]
\begin{centering}
\begin{tabular}{|c|c|c|c|c|c|}
\hline 
\multicolumn{1}{|c||}{$\lambda$} & \multicolumn{1}{c||}{$f_{0}$} & \multicolumn{1}{c||}{$f_{1}$} & \multicolumn{1}{c||}{$f_{2}$} & \multicolumn{1}{c||}{$f_{3}$} & $f_{4}$\tabularnewline
\hline 
\hline 
$-\infty$ & $+\infty$ & $-\infty$ & $+\infty$ & $+\infty$ & $\lambda_{0}-D\left(n\right)$\tabularnewline
\hline 
1 & $\lambda_{0}-C\left(n\right)$ & $-\left(n-4\right)\left(\frac{n-1}{n-2}\right)^{n-1}$ & $\frac{4\left(n-2\right)^{2}}{n^{2}\left(n-3\right)}\left(C\left(n\right)+1\right)-\lambda_{0}-1$ & $\left(\lambda_{0}+1\right)$$\frac{n^{2}\left(n-4\right)}{4\left(n-2\right)}$ & $\lambda_{0}-D\left(n\right)$\tabularnewline
\hline 
\end{tabular}
\par\end{centering}

\protect\caption{For even $n$, the values of the Sturm functions at the endpoints
of the unitary interval of $\lambda$.}
\end{table}
Table 1 summarizes the results. Number of roots depends on the zeros
of the expressions in the last row in Table I:
\begin{equation}
\lambda_{0}=C\left(n\right),\qquad\lambda_{0}=\frac{4\left(n-2\right)^{2}}{n^{2}\left(n-3\right)}\left(C\left(n\right)+1\right)-1,\qquad\lambda_{0}=-1,\qquad\lambda_{0}=D\left(n\right),
\end{equation}
where we have defined 
\begin{equation}
D\left(n\right)\equiv\left(\frac{2}{n}\right)^{n}\left(n-1\right)^{n-1}-1.
\end{equation}
To scan the viable interval of $\lambda_{0}$ that is $\lambda_{0}\in\left(-\infty,C\left(n\right)\right)-\left\{ -1\right\} $
the order of these zeros is important. Let us prove the relations
\begin{equation}
-1<C\left(n\right)\le D\left(n\right).\label{eq:C<D}
\end{equation}
 First, let us show that 
\begin{equation}
C\left(n\right)\le D\left(n\right),\label{eq:Inequality_of_bounds-1}
\end{equation}
where the equality is satisfied for $n=4$. For $n>4$, showing this
inequality boils down to proving the following inequality
\begin{equation}
\left(\frac{2}{n}\right)^{n}>\left(n-3\right)\left(\frac{1}{n-2}\right)^{n},
\end{equation}
since
\begin{equation}
\left(\frac{2}{n}\right)^{n}>\left(\frac{1}{n-2}\right)^{n-1}>\left(n-3\right)\left(\frac{1}{n-2}\right)^{n},
\end{equation}
it suffices to show

\begin{equation}
\frac{n}{2}\left(\frac{2}{n}\right)^{n-1}>\left(\frac{2}{n}\right)^{n-1}>\left(\frac{1}{n-2}\right)^{n-1},
\end{equation}
since $n-2>\frac{n}{2}$, (\ref{eq:Inequality_of_bounds-1}) follows.
Then, since
\begin{align}
\frac{4\left(n-2\right)^{2}}{n^{2}\left(n-3\right)}\left(C\left(n\right)+1\right)-1 & \le C\left(n\right)\\
-\frac{\left(n-4\right)\left(n\left(n-3\right)+4\right)}{n^{2}\left(n-3\right)} & \le0,
\end{align}
where, clearly, equality holds for $n=4$ and inequality holds for
$n>4$. So,
\[
-1<\frac{4\left(n-2\right)^{2}}{n^{2}\left(n-3\right)}\left(C\left(n\right)+1\right)-1\le C\left(n\right)\le D\left(n\right).
\]
In four dimensions, $C\left(4\right)$ and $\left(\frac{2}{n}\right)^{n}\left(n-1\right)^{n-1}-1$
are both equal to $\frac{11}{16}$. This is an important observation
since as we discuss below $D\left(n\right)$ represents the bound
on the existence of roots, that is for $\lambda_{0}\le D\left(n\right)$
there are two roots for the vacuum field equation. For $n=4$, so
if there exist two vacua of the theory, then one of them should be
the viable unitary vacuum. On the other hand, for even dimensional
theories beyond four dimensions, there may be two roots for the vacuum
field equation, but none of them would be in the unitary interval
for $\lambda_{0}$ values $C\left(n\right)<\lambda_{0}<D\left(n\right)$. 

In the tables below, for all values of $\lambda_{0}$, the number
of roots in the unitarity interval $\lambda<1$ is investigated. As
a result, for each value of $\lambda_{0}$ in the interval $\lambda_{0}<C\left(n\right)$,
there is one and only one root for the vacuum equation in the unitary
interval of $\lambda$, and for $\lambda_{0}>C\left(n\right)$, it
is not possible to have a $\lambda$value in the unitary interval.
The following tables from Table-II to Table-IV depict the sign changes
of the Sturm chain for various $\lambda_{0}$ intervals proving the
uniqueness of the real solution $\lambda$.

\begin{table}[H]
\begin{centering}
\begin{tabular}{|c|c|c|c|c|c|c|}
\hline 
$\lambda$ & $f_{0}$ & $f_{1}$ & $f_{2}$ & $f_{3}$ & $f_{4}$ & \# of sign changes\tabularnewline
\hline 
\hline 
$-\infty$ & + & - & + & + & - & 3\tabularnewline
\hline 
1 & - & - & + & - & - & 2\tabularnewline
\hline 
\end{tabular}
\par\end{centering}

\protect\caption{$\lambda_{0}<-1$ case yielding one real root in $\lambda\in\left(-\infty,1\right)$
interval.}
\end{table}

\begin{table}[H]
\begin{centering}
\begin{tabular}{|c|c|c|c|c|c|c|}
\hline 
$\lambda$ & $f_{0}$ & $f_{1}$ & $f_{2}$ & $f_{3}$ & $f_{4}$ & \# of sign changes\tabularnewline
\hline 
\hline 
$-\infty$ & + & - & + & + & - & 3\tabularnewline
\hline 
1 & - & - & + & + & - & 2\tabularnewline
\hline 
\end{tabular}
\par\end{centering}

\protect\caption{$-1<\lambda_{0}<\frac{4\left(n-2\right)^{2}}{n^{2}\left(n-3\right)}\left(C\left(n\right)+1\right)-1$
case yielding one real root in $\lambda\in\left(-\infty,1\right)$
interval.}
\end{table}

\begin{table}[H]
\begin{centering}
\begin{tabular}{|c|c|c|c|c|c|c|}
\hline 
$\lambda$ & $f_{0}$ & $f_{1}$ & $f_{2}$ & $f_{3}$ & $f_{4}$ & \# of sign changes\tabularnewline
\hline 
\hline 
$-\infty$ & + & - & + & + & - & 3\tabularnewline
\hline 
1 & - & - & - & + & - & 2\tabularnewline
\hline 
\end{tabular}
\par\end{centering}

\protect\caption{$\frac{4\left(n-2\right)^{2}}{n^{2}\left(n-3\right)}\left(C\left(n\right)+1\right)-1<\lambda_{0}<C\left(n\right)$
case yielding one real root in $\lambda\in\left(-\infty,1\right)$
interval.}
\end{table}
The following two tables Table-V and Table-VI show that for the nonunitary
interval of $\lambda_{0}$ there is no real solution.

\begin{table}[H]
\begin{centering}
\begin{tabular}{|c|c|c|c|c|c|c|}
\hline 
$\lambda$ & $f_{0}$ & $f_{1}$ & $f_{2}$ & $f_{3}$ & $f_{4}$ & \# of sign changes\tabularnewline
\hline 
\hline 
$-\infty$ & + & - & + & + & - & 3\tabularnewline
\hline 
1 & + & - & - & + & - & 3\tabularnewline
\hline 
\end{tabular}
\par\end{centering}

\protect\caption{$C\left(n\right)<\lambda_{0}<D\left(n\right)$ case yielding no real
root in $\lambda\in\left(-\infty,1\right)$ interval.}
\end{table}

\begin{table}[H]
\begin{centering}
\begin{tabular}{|c|c|c|c|c|c|c|}
\hline 
$\lambda$ & $f_{0}$ & $f_{1}$ & $f_{2}$ & $f_{3}$ & $f_{4}$ & \# of sign changes\tabularnewline
\hline 
\hline 
$-\infty$ & + & - & + & + & + & 2\tabularnewline
\hline 
1 & + & - & - & + & + & 2\tabularnewline
\hline 
\end{tabular}
\par\end{centering}

\protect\caption{$D\left(n\right)<\lambda_{0}$ case yielding no real root in $\lambda\in\left(-\infty,1\right)$
interval.}
\end{table}

To get a sort of intuitive feeling, this analysis can be done graphically
for each $n$ and a specific viable $\lambda_{0}$: As an example,
see Fig-1. One can show that for $\lambda_{0}<C\left(n\right)$ ,
there is one and only one $\lambda$ value in $\lambda<1$ by investigating
the asymptotes and extrema of $f\left(\lambda\right)$ whose graph
is given in Fig.~1.

\begin{figure}[H]
\begin{centering}
\includegraphics[scale=0.8]{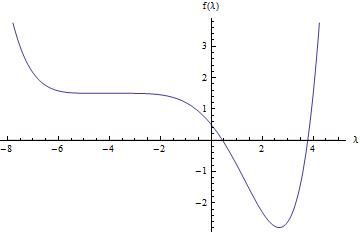}
\par\end{centering}

\protect\caption{As an example, we chose $n=6$, $\lambda_{0}=0.5$, but the results
are generic.}
\end{figure}

Note that this graph is representative of the generic shape of the
$f\left(y\right)$ function for any even $n$ and for any $\lambda_{0}$.
This can be seen as follows: $\underset{y\rightarrow\pm\infty}{\lim}f\left(y\right)\rightarrow+\infty$,
the inflection point $y_{{\rm inf}}=0$ satisfies $y_{{\rm inf}}<y_{{\rm min}}=\frac{2\left(n-1\right)}{n}$,
finally the value of the function at the inflection point is always
larger than the value of the function at $y_{{\rm min}}$, namely
\begin{equation}
a=f\left(0\right)>f\left(\frac{2\left(n-1\right)}{n}\right)=-\left(\frac{2}{n}\right)^{n}\left(n-1\right)^{n-1}+a.
\end{equation}
Therefore, the equation $f\left(y\right)=0$ has real root(s) if and
only if
\begin{equation}
f\left(\frac{2\left(n-1\right)}{n}\right)\le0,\label{eq:Root_condition-1}
\end{equation}
when the bound is saturated there is a unique root. This condition
requires an upper bound on $\lambda_{0}$ as 
\begin{equation}
\lambda_{0}\le D\left(n\right)=\left(\frac{2}{n}\right)^{n}\left(n-1\right)^{n-1}-1.
\end{equation}
One should compare this bound, coming from the existence of a vacuum,
to the bound on $\lambda_{0}$, coming from the unitarity of the theory
(\ref{eq:Unitarity_bound-1}). 

\emph{The conclusion of the above analysis is that for even dimensions,
there is a unique viable vacuum, $\lambda$, in the region $-\infty<\lambda<1$
and $\lambda\ne2-n$ given that $-\infty<\lambda_{0}<C\left(n\right)$
and $\lambda_{0}\ne-1$.}\textcolor{red}{{} }

\subsubsection{Odd dimensions: $n=5+2k$, $k=0,1,\dots$}

For unitarity $\tilde{\kappa}>0$ must be satisfied which requires
$2-n<\lambda<1$ in odd dimensions as is clear from (\ref{eq:kappa_tilde_of_the_theory}).
Therefore, for odd dimensions, the relevant interval is $y\in\left(0,\frac{n-1}{n-2}\right)$.
There is a corresponding viable interval of $\lambda_{0}$ since $\lambda_{0}=\lambda_{0}\left(y\right)$,
that is (\ref{eq:lamda0_in_lambda-1}), is a monotonically increasing
function of $y$ as $d\lambda_{0}/dy$, that is (\ref{eq:Derv_of_lambda0-1}),
is positive in the unitarity region due to positivity of $y$ and
the same reasoning in the even-dimensional case follows. The lower
bound of $y$ yields a lower bound for $\lambda_{0}$, so unitarity
region of $\lambda$ corresponds to $-1<\lambda_{0}<C\left(n\right)$.
In Table 7, the values of the Sturm functions are given at the endpoints
of the unitary interval of $\lambda$.

\begin{table}[H]
\begin{centering}
\begin{tabular}{|c|c|c|c|c|c|}
\hline 
\multicolumn{1}{|c||}{$\lambda$} & \multicolumn{1}{c||}{$f_{0}$} & \multicolumn{1}{c||}{$f_{1}$} & \multicolumn{1}{c||}{$f_{2}$} & \multicolumn{1}{c||}{$f_{3}$} & $f_{4}$\tabularnewline
\hline 
\hline 
$2-n$ & $\lambda_{0}+1$ & $0$ & $-\lambda_{0}-1$ & $\left(\lambda_{0}+1\right)\frac{n^{2}}{2}$ & $\lambda_{0}-D\left(n\right)$\tabularnewline
\hline 
1 & $\lambda_{0}-C\left(n\right)$ & $-\left(n-4\right)\left(\frac{n-1}{n-2}\right)^{n-1}$ & $\frac{4\left(n-2\right)^{2}}{n^{2}\left(n-3\right)}\left(C\left(n\right)+1\right)-\lambda_{0}-1$ & $\left(\lambda_{0}+1\right)$$\frac{n^{2}\left(n-4\right)}{4\left(n-2\right)}$ & $\lambda_{0}-D\left(n\right)$\tabularnewline
\hline 
\end{tabular}
\par\end{centering}

\protect\caption{For odd $n$, the values of the Sturm functions at the endpoints of
the unitary interval of $\lambda$.}
\end{table}
The number of roots depends on the zeros of the expressions in the
second and third rows of the Table VII. In the second row, the only
root is $\lambda_{0}=-1$ which was already a root of the third row.
The roots in the third row were investigated in the even dimensional
case and they can be ordered as
\begin{equation}
-1<\frac{4\left(n-2\right)^{2}}{n^{2}\left(n-3\right)}\Bigg(C\left(n\right)+1\Bigg)-1\le C\left(n\right)\le D\left(n\right).
\end{equation}
In the tables below, for all values of $\lambda_{0}$, the number
of roots in the unitarity interval $2-n<\lambda<1$ is investigated.
As a result, for each value of $\lambda_{0}$ in the interval $-1<\lambda_{0}<C\left(n\right)$,
there is one and only one root for the vacuum equation in the unitary
interval of $\lambda$, and for $\lambda_{0}<-1$ and $\lambda_{0}>C\left(n\right)$,
it is not possible to have a $\lambda$ value in the unitary interval.

\begin{table}[H]
\begin{centering}
\begin{tabular}{|c|c|c|c|c|c|c|}
\hline 
$\lambda$ & $f_{0}$ & $f_{1}$ & $f_{2}$ & $f_{3}$ & $f_{4}$ & \# of sign changes\tabularnewline
\hline 
\hline 
$2-n$ & - & 0 & + & - & - & 2\tabularnewline
\hline 
1 & - & - & + & - & - & 2\tabularnewline
\hline 
\end{tabular}
\par\end{centering}

\protect\caption{$\lambda_{0}<-1$ case yielding no real root in $\lambda\in\left(2-n,1\right)$
interval.}
\end{table}

\begin{table}[H]
\begin{centering}
\begin{tabular}{|c|c|c|c|c|c|c|}
\hline 
$\lambda$ & $f_{0}$ & $f_{1}$ & $f_{2}$ & $f_{3}$ & $f_{4}$ & \# of sign changes\tabularnewline
\hline 
\hline 
$2-n$ & + & 0 & - & + & - & 3\tabularnewline
\hline 
1 & - & - & + & + & - & 2\tabularnewline
\hline 
\end{tabular}
\par\end{centering}

\protect\caption{$-1<\lambda_{0}<\frac{4\left(n-2\right)^{2}}{n^{2}\left(n-3\right)}\left(C\left(n\right)+1\right)-1$
case yielding one real root in $\lambda\in\left(2-n,1\right)$ interval.}
\end{table}

\begin{table}[H]
\begin{centering}
\begin{tabular}{|c|c|c|c|c|c|c|}
\hline 
$\lambda$ & $f_{0}$ & $f_{1}$ & $f_{2}$ & $f_{3}$ & $f_{4}$ & \# of sign changes\tabularnewline
\hline 
\hline 
$2-n$ & + & 0 & - & + & - & 3\tabularnewline
\hline 
1 & - & - & - & + & - & 2\tabularnewline
\hline 
\end{tabular}
\par\end{centering}

\protect\caption{$\frac{4\left(n-2\right)^{2}}{n^{2}\left(n-3\right)}\left(C\left(n\right)+1\right)-1<\lambda_{0}<C\left(n\right)$
case yielding one real root in $\lambda\in\left(2-n,1\right)$ interval.}
\end{table}

\begin{table}[H]
\begin{centering}
\begin{tabular}{|c|c|c|c|c|c|c|}
\hline 
$\lambda$ & $f_{0}$ & $f_{1}$ & $f_{2}$ & $f_{3}$ & $f_{4}$ & \# of sign changes\tabularnewline
\hline 
\hline 
$2-n$ & + & 0 & - & + & - & 3\tabularnewline
\hline 
1 & + & - & - & + & - & 3\tabularnewline
\hline 
\end{tabular}
\par\end{centering}

\protect\caption{$C\left(n\right)<\lambda_{0}<D\left(n\right)$ case yielding no real
root in $\lambda\in\left(2-n,1\right)$ interval.}
\end{table}

\begin{table}[H]
\begin{centering}
\begin{tabular}{|c|c|c|c|c|c|c|}
\hline 
$\lambda$ & $f_{0}$ & $f_{1}$ & $f_{2}$ & $f_{3}$ & $f_{4}$ & \# of sign changes\tabularnewline
\hline 
\hline 
$2-n$ & + & 0 & - & + & + & 2\tabularnewline
\hline 
1 & + & - & - & + & + & 2\tabularnewline
\hline 
\end{tabular}
\par\end{centering}

\protect\caption{$D\left(n\right)<\lambda_{0}$ case yielding no real root in $\lambda\in\left(2-n,1\right)$
interval.}
\end{table}

Again, for given $n$ and $\lambda_{0}$ this analysis can be done
graphically by investigating the asymptotes and extrema of $f\left(\lambda\right)$
whose graph is given in Fig. 2.

\begin{figure}[H]
\begin{centering}
\includegraphics[scale=0.2]{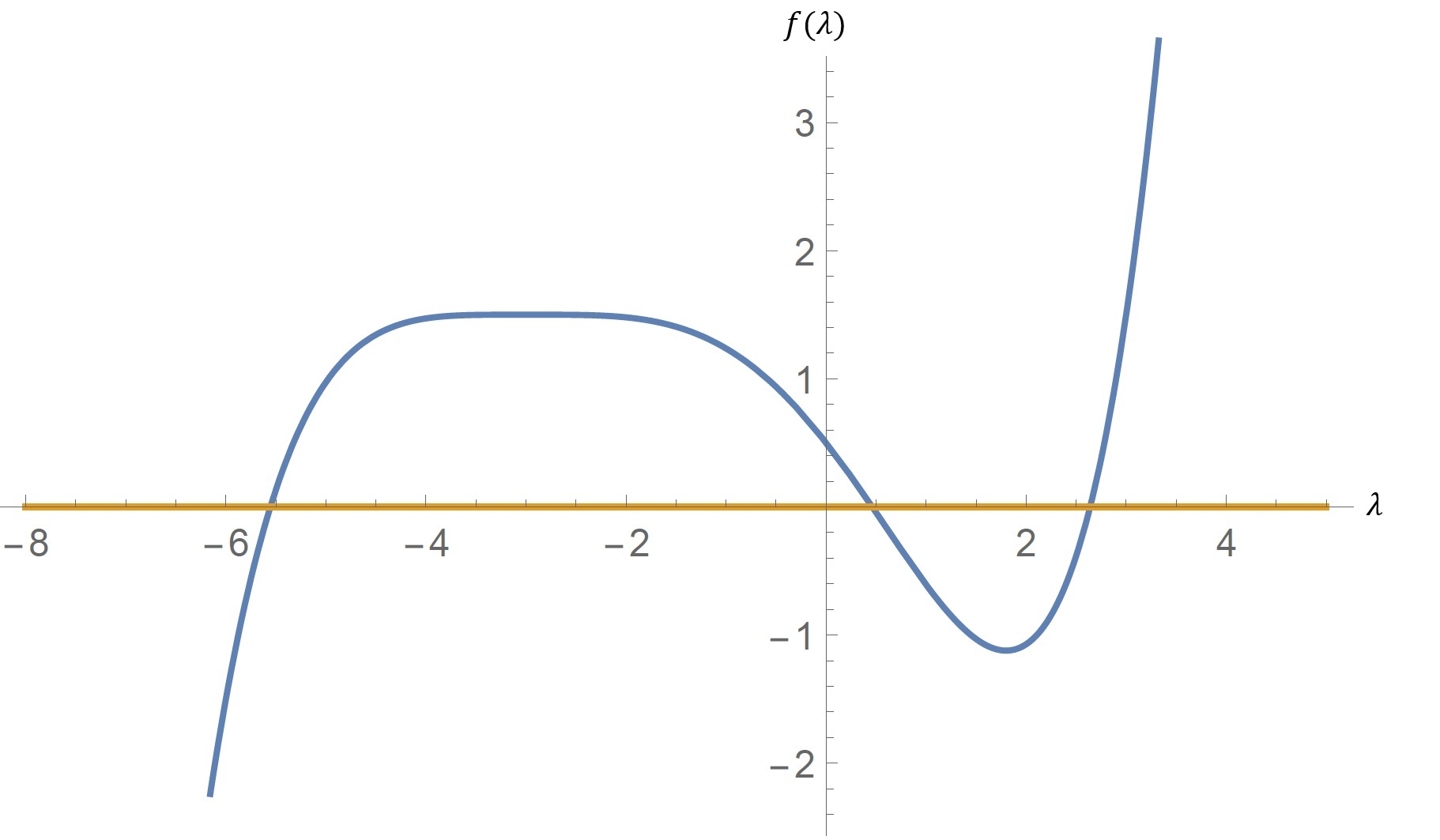}
\par\end{centering}

\protect\caption{$n=5$, $\lambda_{0}=0.5$}
\end{figure}

Note that this graph is representative of the generic shape of $f\left(y\right)$
function for any odd $n$ and for any $\lambda_{0}$. This can be
seen as follows: $\underset{y\rightarrow\pm\infty}{\lim}f\left(y\right)\rightarrow\pm\infty$,
the inflection point $y_{{\rm inf}}=0$ satisfies $y_{{\rm inf}}<y_{{\rm min}}=\frac{2\left(n-1\right)}{n}$,
finally the value of the function at the inflection point is always
larger than the value of the function at $y_{{\rm min}}$, namely
\[
a=f\left(0\right)>f\left(\frac{2\left(n-1\right)}{n}\right)=-\left(\frac{2}{n}\right)^{n}\left(n-1\right)^{n-1}+a.
\]
Therefore, the equation $f\left(y\right)=0$ has one real root for
either $f\left(y_{{\rm inf}}\right)<0$ or $f\left(y_{{\rm min}}\right)>0$,
and there are three real roots if $f\left(y_{{\rm inf}}\right)>0$
and $f\left(y_{{\rm min}}\right)<0$. In addition, there are two real
roots for either $f\left(y_{{\rm inf}}\right)=0$ or $f\left(y_{{\rm min}}\right)=0$.
For the one real root case, $\lambda_{0}$ should satisfy either
\[
f\left(0\right)=a<0\Rightarrow\lambda_{0}<-1,
\]
or
\[
f\left(\frac{2\left(n-1\right)}{n}\right)=-\left(\frac{2}{n}\right)^{n}\left(n-1\right)^{n-1}+a>0\Rightarrow\lambda_{0}>\left(\frac{2}{n}\right)^{n}\left(n-1\right)^{n-1}-1\Rightarrow\lambda_{0}>D\left(n\right).
\]
But, since $\lambda_{0}$ should be in the interval $-1<\lambda_{0}<C\left(n\right)$
to have a unitary root, these conditions cannot be satisfied for a
unitary theory since $C\left(n\right)\le D\left(n\right)$. So, for
a unitary theory in odd dimensions, there cannot be one real root
only. Moving to the case of three real roots which is possible if
$-1<\lambda_{0}<D\left(n\right)$ is satisfied. Therefore, if the
theory is unitary, that is if$-1<\lambda_{0}<C\left(n\right)$, then
there are always three real roots since $C\left(n\right)\le D\left(n\right)$,
and as $\lambda$ and $\lambda_{0}$ are related to each other with
a 1-1 correspondence in the unitary interval, there is always a unique
viable vacuum and two nonunitary vacua. Lastly, for the case of two
real roots, either $\lambda_{0}=-1$, which is not possible, or $\lambda_{0}=D\left(n\right)$
which cannot be satisfied for an odd dimensional unitary theory. Hence,
the case of two real roots does not yield a viable theory.

\emph{The conclusion of the above analysis is that for odd dimensions,
there is a unique viable vacuum, $\lambda$, in the region $2-n<\lambda<1$
given that $-1<\lambda_{0}<C\left(n\right)$.}


\begin{thebibliography}{References}
\bibitem{Gullu-4BI} I.~Gullu, T.~C.~Sisman and B.~Tekin, ``\emph{Born-Infeld
Gravity with a Massless Graviton in Four Dimensions},'' Phys.\ Rev.\ D
\textbf{91}, no. 4, 044007 (2015).

\bibitem{Deser_Gibbons} S.~Deser and G.~W.~Gibbons, \emph{``Born-Infeld-Einstein
actions?,''} Class.\ Quant.\ Grav.\ \textbf{15}, L35 (1998). 

\bibitem{Gullu-BINMG} I.~Gullu, T.~C.~Sisman, B.~Tekin, \emph{``Born-Infeld
extension of new massive gravity,''} Class.\ Quant.\ Grav.\ \textbf{27},
162001 (2010).

\bibitem{Gullu-cfunc} I.~Gullu, T.~C.~Sisman and B.~Tekin, \emph{``c-functions
in the Born-Infeld extended New Massive Gravity,''} Phys.\ Rev.\ D
\textbf{82}, 024032 (2010).

\bibitem{Stelle} K.~S.~Stelle, \emph{``Renormalization of Higher
Derivative Quantum Gravity,''} Phys.\ Rev.\ D\textbf{ 16}, 953
(1977).

\bibitem{BIuniD-short} I.~Gullu, T.~C.~Sisman and B.~Tekin, to
appear soon.

\bibitem{Boulware-String} D.~G.~Boulware and S.~Deser, ``String
Generated Gravity Models,'' Phys.\ Rev.\ Lett.\ \textbf{55}, 2656
(1985).

\bibitem{Gullu-UniBI} I.~Gullu, T.~C.~Sisman and B.~Tekin, \emph{``Unitarity
analysis of general Born-Infeld gravity theories,''} Phys.\ Rev.\ D
\textbf{82}, 124023 (2010).

\bibitem{Sisman-Thesis} T.~C.~Sisman, ``\emph{Born-Infeld gravity
theories in $D$-dimensions,}'' PhD thesis, METU, (2012).

\bibitem{Hindawi} A.~Hindawi, B.~A.~Ovrut and D.~Waldram, \emph{``Nontrivial
vacua in higher derivative gravitation,''} Phys.\ Rev.\ D \textbf{53},
5597 (1996).

\bibitem{Sisman-AllUniD} T.~C.~Sisman, I.~Gullu and B.~Tekin,
\emph{``All unitary cubic curvature gravities in D dimensions,''}
Class.\ Quant.\ Grav.\ \textbf{28}, 195004 (2011).

\bibitem{Gullu-AllUni3D} I.~Gullu, T.~C.~Sisman, B.~Tekin, \emph{``All
Bulk and Boundary Unitary Cubic Curvature Theories in Three Dimensions,''}
Phys.\ Rev.\ D\textbf{ 83}, 024033 (2011).

\bibitem{Senturk} C.~Senturk, T.~C.~Sisman and B.~Tekin, \emph{``Energy
and Angular Momentum in Generic F(Riemann) Theories,''} Phys.\ Rev.\ D
\textbf{86}, 124030 (2012).

\bibitem{Abbott-Deser} L.~F.~Abbott and S.~Deser, ``\emph{Stability
of Gravity with a Cosmological Constant,}'' Nucl. Phys. B \textbf{195},
76 (1982).

\bibitem{Deser_Tekin-PRL} S.~Deser and B.~Tekin, \emph{``Gravitational
energy in quadratic curvature gravities,''} Phys.\ Rev.\ Lett.\ \textbf{89},
101101 (2002).

\bibitem{Deser_Tekin} S.~Deser and B.~Tekin, \emph{``Energy in
generic higher curvature gravity theories,''} Phys.\ Rev.\ D\textbf{
67}, 084009 (2003).

\bibitem{Emparan} R.~Emparan, D.~Grumiller and K~Tanabe, ``\emph{Large-D
gravity and low-D strings,}'' Phys.\ Rev.\ Lett.\ \textbf{110}
, 251102 (2013).

\bibitem{Eddington} A.~Eddington, \emph{The Mathematical Theory
of General Relativity} (Cambridge University Press, Cambridge, England,
1924).

\bibitem{BI} M.~Born and L.~Infeld, \emph{``Foundations of the
new field theory,''} Proc.\ Roy.\ Soc.\ Lond.\ A \textbf{144},
425 (1934).

\bibitem{Banados_Eddington} M.~Banados and P.~G.~Ferreira, \emph{``Eddington's
theory of gravity and its progeny,''} Phys.~Rev.~Lett. \textbf{105},
011101 (2010).

\bibitem{Delsate_Steinhoff} T.~Delsate and J.~Steinhoff, \emph{``New
insights on the matter-gravity coupling paradigm,''} Phys.~Rev.~Lett.
\textbf{109}, 021101 (2012).

\bibitem{Fiorini} F.~Fiorini, \emph{``Nonsingular Promises from
Born-Infeld Gravity,}'' Phys.~Rev.~Lett. \textbf{111}, 041104 (2013).

\bibitem{Lavinia-Infrared}  J.~B.~Jiménez, L.~Heisenberg and G.~J.~Olmo,
``\emph{Infrared lessons for ultraviolet gravity: the case of massive
gravity and Born-Infeld,}'' JCAP \textbf{1411}, 004 (2014).

\bibitem{Lavinia-cascading} J.~B.~Jiménez, L.~Heisenberg and G.~J.~Olmo
and C.~Ringeval, ``\emph{Cascading dust inflation in Born-Infeld
gravity,}'' arXiv:1509.01188 {[}gr-qc{]}.

\bibitem{Cadabra-1} K.~Peeters, ``\emph{Cadabra: a field-theory
motivated symbolic computer algebra system}'', Comput.\ Phys.\ Commun.,
\textbf{176} , 550 (2007). 

\bibitem{Cadabra-2} K.~Peeters, ``\emph{Introducing Cadabra: A
symbolic computer algebra system for field theory problems}'', arXiv:hep-th/
0701238.

\bibitem{Conk} N.~B.~Conkwright, ``\emph{Introduction to the Theory
of Equations},'' Ginn and Company, Boston, MA, (1957).\end{thebibliography}
\end{document}